\documentclass[usegraphicx,usenatbib,letterpaper]{emulateapj}

\usepackage{ulem}
\usepackage{color}
\usepackage{graphics,graphicx}
\usepackage{times} 
\usepackage{amssymb}
\usepackage{amsmath}
\usepackage{natbib}
\usepackage{url}
\usepackage{multirow}
\usepackage{ctable}
\usepackage{threeparttable}
\newif\ifAMStwofonts
\AMStwofontstrue
\def\rxte{{\it RXTE}}

\def\ginga{{\it Ginga}}

\def\chandra{{\it Chandra}}

\def\integral{{\it INTEGRAL~\/}}

\def\epicmos1{{EPIC-MOS1}}
\def\epicmos2{{EPIC-MOS2}}
\def\epicmos{{EPIC-MOS}}

\def\nustar{{\it NuSTAR}}

\def\deg{$^{\circ}$}

\def\H0{{\rm ~km~s^{-1}~Mpc^{-1}}}

\def\kev{\hbox{\rm keV}}

\def\ctps{\hbox{$\rm\thinspace ct~s^{-1}$}}

\def\ergpcmsqps{\hbox{$\rm\thinspace erg~cm^{-2}~s^{-1}$}}

\def\ergps{\hbox{erg~s$^{-1}$}}
\def\ergcmps{\hbox{\rm erg~cm~s$^{-1}$}}

\def\ledd{$L_{\rm{E}}$}

\def\msun{\hbox{$\rm M_{\odot}$}}

\def\chisq{{$\chi^{2}$}}

\def\xspec{\hbox{\small XSPEC}}
\def\xspecv{\hbox{\small XSPEC}\, v12.6.0f}

\def\nustardas{\rm {\small NUSTARDAS}}

\def\addascaspec{\hbox{\rm{\small ADDASCASPEC~\/}}}

\def\addascaspec{\hbox{\rm{\small ADDASCASPEC}}}
\def\nupipeline{\rm{\small NUPIPELINE}}
\def\nuproducts{\rm{\small NUPRODUCTS}}

\def\xstar{\hbox{\rm{\small XSTAR}}}

\def\grid25{\hbox{\rm{\small GRID25}}}

\def\tbabs{\rm{\small TBABS}}

\def\diskbb{\rm{\small DISKBB}}

\def\reflionx{\rm{\small REFLIONX}}
\def\xillver{\rm{\small XILLVER}}
\def\relxill{\rm{\small RELXILL}}
\def\relxilllp{\rm{\small RELXILLLP}}

\def\relconv{\rm{\small RELCONV}}
\def\relline{\rm{\small RELLINE}}

\def\fexxv{\hbox{\rm Fe\,{\small XXV}}}
\def\fexxvi{\hbox{\rm Fe\,{\small XXVI}}}

\def\eg{{\it e.g.}}

\def\ie{{\it i.e.}}

\def\la{\mathrel{\hbox{\rlap{\hbox{\lower4pt\hbox{$\sim$}}}{\raise2pt\hbox{$<$}}}}}
\def\ga{\mathrel{\hbox{\rlap{\hbox{\lower4pt\hbox{$\sim$}}}{\raise2pt\hbox{$>$}}}}}

\def\d25{D$_{25}$}
\def\nh{{$N_{\rm H}$}}

\def\.25{0.25 keV\thinspace}

\def\rg{$r_{\rm{G}}$}
\def\rh{$r_{\rm{H}}$}
\def\rin{$r_{\rm{in}}$}
\def\rout{$r_{\rm{out}}$}

\def\risco{$r_{\rm{ISCO}}$}

\def\v404{\rm V404\,Cyg}
\def\hredge{$R_{\rm{edge}}$}
\def\felim{0.9}
\def\ilim{45}

\shorttitle{Relativistic Reflection in \v404}
\shortauthors{D.~J. Walton et al.}

\begin{document}

\title{Living on a Flare: Relativistic Reflection in V404 Cyg Observed by \textit{NuSTAR} During its Summer 2015 Outburst}

\author{D. J. Walton\altaffilmark{1,2,3},
K. Mooley\altaffilmark{4},
A. L. King\altaffilmark{5},
J. A. Tomsick\altaffilmark{6},
J. M. Miller\altaffilmark{7},
T. Dauser\altaffilmark{8},
J. A. Garc\'ia\altaffilmark{2,8,9},
M. Bachetti\altaffilmark{10},\\
M. Brightman\altaffilmark{2},
A. C. Fabian\altaffilmark{3},
%R. P. Fender\altaffilmark{4},
K. Forster\altaffilmark{2},
F. F\"{u}rst\altaffilmark{2},
P. Gandhi\altaffilmark{11},
B. W. Grefenstette\altaffilmark{2},
F. A. Harrison\altaffilmark{2},\\
K. K. Madsen\altaffilmark{2},
D. L. Meier\altaffilmark{1,2},
M. J. Middleton\altaffilmark{11},
L. Natalucci\altaffilmark{12},
F. Rahoui\altaffilmark{13,14},
V. Rana\altaffilmark{2},
D. Stern\altaffilmark{1}
}
\affil{ \\
$^{1}$ Jet Propulsion Laboratory, California Institute of Technology, Pasadena, CA 91109, USA \\
$^{2}$ Space Radiation Laboratory, California Institute of Technology, Pasadena, CA 91125, USA \\
$^{3}$ Institute of Astronomy, University of Cambridge, Madingley Road, Cambridge CB3 0HA, UK \\
$^{4}$ Department of Physics, Oxford University, Denys Wilkinson Building, Keble Road, Oxford OX1 3RH, UK \\
$^{5}$ Einstein Fellow, Department of Physics, Stanford University, 382 Via Pueblo Mall, Stanford, CA 94305, USA \\
$^{6}$ Space Sciences Laboratory, University of California, Berkeley, CA 94720, USA \\
$^{7}$ Department of Astronomy, University of Michigan, 1085 S. University Ave., Ann Arbor, MI, 49109-1107, USA \\
$^{8}$ ECAP-Erlangen Centre for Astroparticle Physics, Sternwartstrasse 7, D-96049 Bamberg, Germany \\
$^{9}$ Harvard-Smithsonian Center for Astrophysics, 60 Garden Street, Cambridge, MA 02138, USA \\
$^{10}$ INAF/Osservatorio Astronomico di Cagliari, via della Scienza 5, I-09047 Selargius (CA), Italy \\
$^{11}$ Department of Physics and Astronomy, University of Southampton, Highfield, Southampton SO17 1BJ, UK \\
$^{12}$ Istituto di Astrofisica e Planetologia Spaziali, INAF, Via Fosso del Cavaliere 100, I-00133 Roma, Italy \\
$^{13}$ European Southern Observatory, K. Schwarzschild-Str. 2, 85748 Garching bei München, Germany \\
$^{14}$ Department of Astronomy, Harvard University, 60 Garden Street, Cambridge, MA 02138, USA \\
}

\begin{abstract}
We present first results from a series of \nustar\ observations of the black hole X-ray
binary \v404\ obtained during its summer 2015 outburst, primarily focusing on
observations during the height of this outburst activity. The \nustar\ data show extreme
variability in both the flux and spectral properties of the source. This is partly driven by
strong and variable line-of-sight absorption, similar to previous outbursts. The latter
stages of this observation are dominated by strong flares, reaching luminosities close
to Eddington. During these flares, the central source appears to be relatively
unobscured and the data show clear evidence for a strong contribution from
relativistic reflection, providing a means to probe the geometry of the innermost
accretion flow. Based on the flare properties, analogy with other Galactic black hole
binaries, and also the simultaneous onset of radio activity, we argue that this intense
X-ray flaring is related to transient jet activity during which the ejected plasma is the
primary source of illumination for the accretion disk. If this is the case, then our
reflection modelling implies that these jets are launched in close proximity to the black
hole (as close as a few gravitational radii), consistent with expectations for jet
launching models that tap either the spin of the central black hole, or the very
innermost accretion disk. Our analysis also allows us to place the first constraints on
the black hole spin for this source, which we find to be $a^* > 0.92$ (99\% statistical
uncertainty, based on an idealized lamppost geometry).
\end{abstract}


\section{Introduction}

V404\,Cygni (hereafter \v404, also known as GS 2023+338) is a well-known,
dynamically confirmed black hole X-ray binary (BHB). The black hole, of mass
9--15\,\msun, is in a 6.5d binary system with a lower mass K-type stellar
companion, from which it accretes via Roche-lobe overflow (\citealt{Casares92nat,
Wagner92, Shahbaz94, Sanwal96, Khargharia10}). Located only $2.39 \pm 0.14$
kpc away (\citealt{MillerJones09}), \v404\ is one of the closest black hole systems
known (\citealt{blackcat, watchdog}).

\begin{figure*}
\hspace*{-0.5cm}
\epsscale{1.1}
\plotone{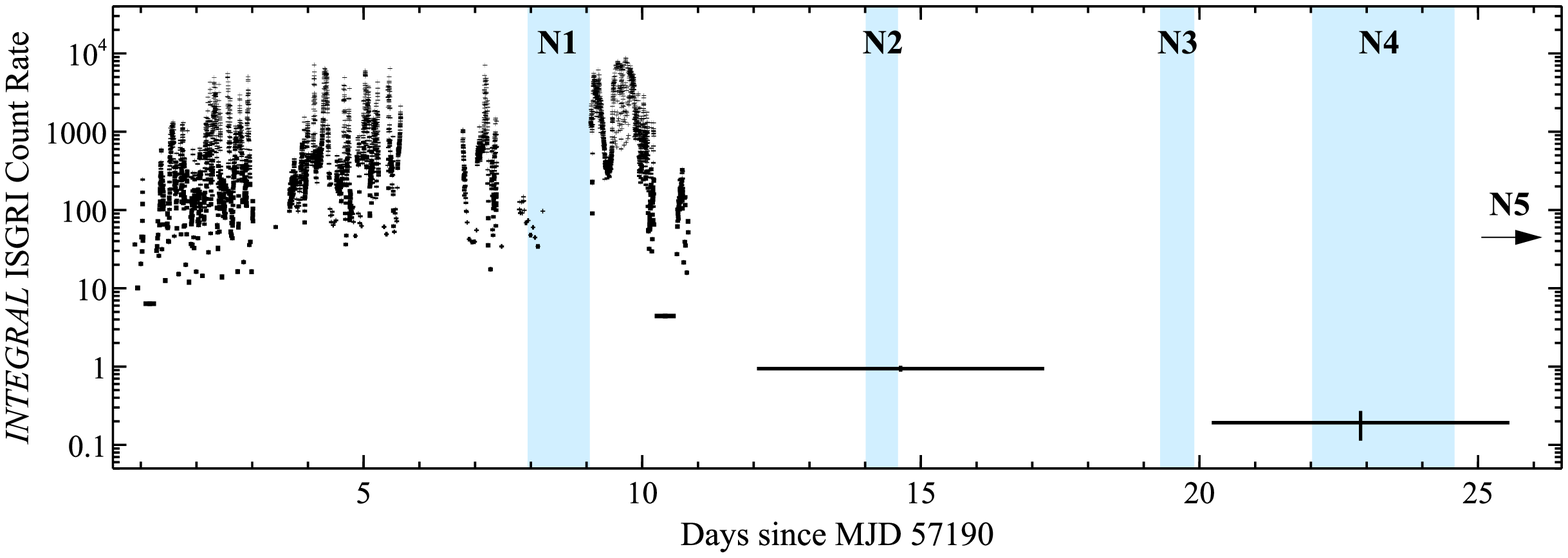}
\caption{Long-term 25--200\,keV X-ray lightcurve for the recent outburst from \v404\
observed with \integral\ (see \citealt{Kuulkers16v404} for details). The first four of our
five \nustar\ observations are indicated with the shaded regions (N1--4; the fifth, N5,
spanned MJD $\sim$57226.35--57227.46); the first caught \v404\ during the height
of its activity, and is the subject of this work, while the following four observations
probed various stages of its decline back to quiescence (Rana et al, \textit{in
preparation}).
}
\vspace{0.3cm}
\label{fig_longlc}
\end{figure*}

As a low-mass X-ray binary (LMXB), \v404\ spends the majority of its time in
quiescence, and has become one of the key targets for studying black holes in
this regime (\eg\ \citealt{Reynolds14quies, Bernardini14, Rana16}). However, as
with other LMXBs, it undergoes intense accretion outbursts, likely related to the
hydrogen ionization instability (see \citealt{Lasota01rev} for a review). Although
these events are rare, during these outbursts \v404\ becomes one of the brightest
X-ray sources in the sky. The X-ray band is vital for studying the accretion flow.
For BHBs, the thermal emission from the accretion disk, the high-energy powerlaw
continuum (likely resulting from Compton up-scattering of the disk emission), and
the disk reflection spectrum (resulting from irradiation of the disk) all contribute to
the broadband X-ray emission (\eg\ \citealt{Zdziarski02, Reis10lhs,
Walton12xrbAGN, Tomsick14}; see \citealt{Done07rev} for a review). The disk
reflection spectrum is particularly critical, as this carries information regarding
both the geometry of the innermost accretion flow (\eg\ \citealt{Wilkins12,
Dauser13}) and the spin of the central black hole (\eg\ \citealt{Miller09,
Reis09spin, Brenneman11, Walton13spin}; see \citealt{Reynolds14rev} and
\citealt{Middleton15rev} for recent reviews). \v404\ is therefore an important
source with which to investigate these accretion phenomena.

However, in some respects, \v404\ is unusual for a black hole LMXB.
Throughout a typical outburst, most sources follow a relatively well-defined
pattern of accretion states (see \citealt{Fender14rev} and \citealt{Belloni16rev}
for recent reviews). Sources rise from quiescence into the hard state, in which
the powerlaw dominates the emission and persistent radio jets are seen. As
the accretion rate continues to increase sources transition into the soft state,
in which the thermal emission from the disk dominates the observed emission.
The radio jets are believed to be quenched in this state, and outflows are typically
seen in the form of winds from the accretion disk instead (\eg\ \citealt{Miller06a,
Neilsen09, Ponti12}, although recent analyses suggest that jets and disk winds
may not necessarily be mutually exclusive, \citealt{Rahoui14, Reynolds15,
Homan16}). Then, as the sources fade, they move back through the hard state,
before finally returning to quiescence. 

\v404\ instead shows much more complexity. Its major 1989 outburst, which first
identified the source as an X-ray binary, was well covered by the \ginga\
observatory (\citealt{Kitamoto89, Terada94, Oosterbroek97, Zycki99a, Zycki99b}).
These observations revealed extreme levels of variability across a wide range of
timescales. In part, this was driven by large variations in the line-of-sight
absorption column, which was often significantly in excess of that seen during
quiescence. Such variations are not typically seen in other black hole LMXBs.
This strong and variable absorption resulted in complex X-ray spectra, making
identification of standard accretion states extremely challenging. In addition,
evidence for X-ray reprocessing from both ionised and neutral material was
observed at varying intervals, further complicating spectral decomposition
(\eg\ \citealt{Zycki99b}).

In the summer of 2015, \v404\ underwent its first major outburst since 1989,
triggering an enormous multi-wavelength observing campaign (\eg\
\citealt{Rodriguez15v404, Natalucci15, Roques15, King15v404,
Jenke16v404, Gandhi16v404, Kimura16nat, MunozDarias16, Motta16v404}, as
well as many other works in preparation). As part of this broadband follow-up
effort, we undertook a series of high-energy X-ray observations with the
\textit{Nuclear Spectroscopic Telescope Array} (\nustar; \citealt{NUSTAR}). Its
unique combination of unprecedented high-energy sensitivity and broad
bandpass (3--79\,\kev) make \nustar\ extremely well suited for disentangling the
contributions from reflection and absorption (as demonstrated, for example, by
the recent broadband work on the active galaxy NGC\,1365; \citealt{Risaliti13nat,
Walton14, Kara15, Rivers15}), and allows detailed, broadband spectroscopy to
be performed on timescales much shorter than previously accessible. Critically
for \v404, \nustar's triggered read-out means it is also well suited to observing
sources with extremely high count-rates (\eg\ \citealt{Miller13grs, Fuerst15,
Parker16, Walton16cyg}), providing clean, high signal-to-noise measurements of
their spectra without suffering from instrumental issues like photon pile-up, etc.

In this work, we present results from our 2015 \nustar\ campaign on \v404,
focusing on observations made at the height of the outburst activity. The paper is
structured as follows: section \ref{sec_red} describes the \nustar\ observations and
our data reduction procedure, sections \ref{sec_time} and \ref{sec_spec} present
our analysis of the temporal and spectral variability exhibited by \v404, and section
\ref{sec_dis} presents a discussion of the results obtained. Finally, we summarize
our main conclusions in section \ref{sec_conc}.

\begin{figure*}
\hspace*{-0.5cm}
\epsscale{1.1}
\plotone{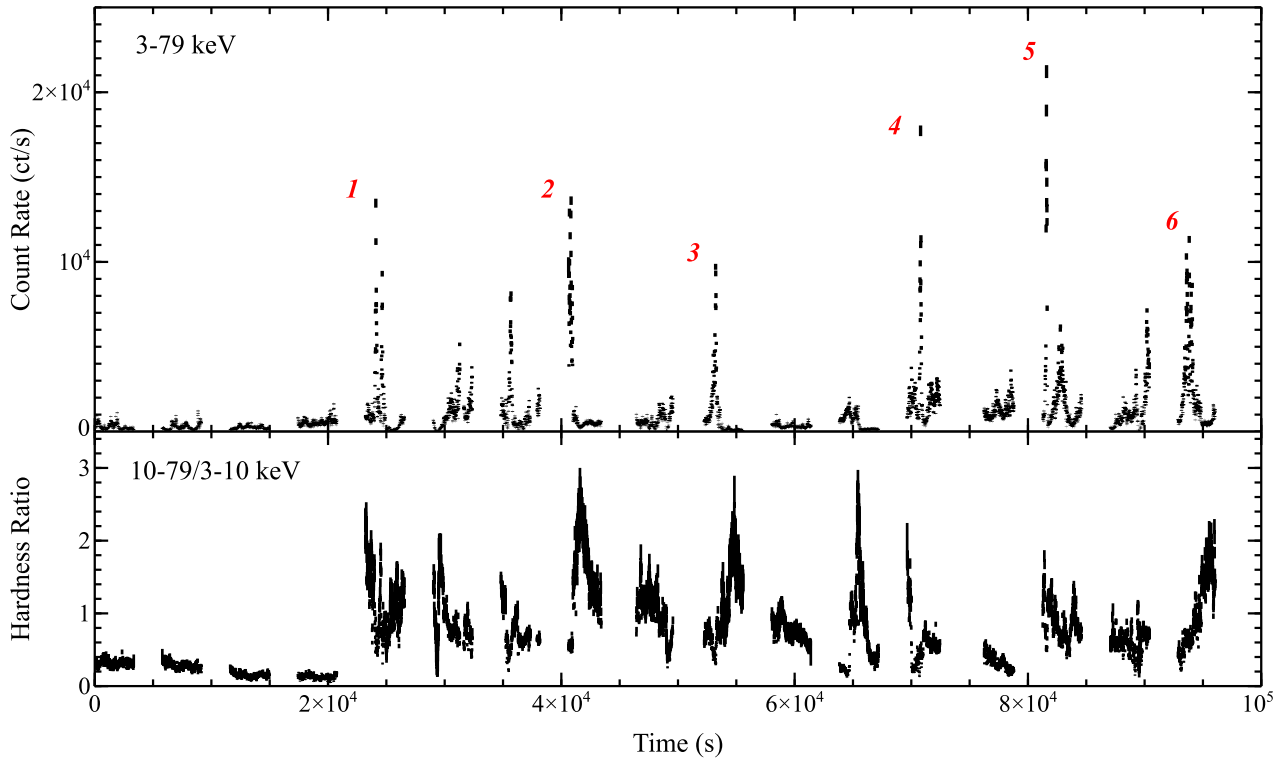}
\caption{\nustar\ lightcurve for the first observation of \v404\ (top panel, 10s bins).
Only the FPMA data are shown for clarity, and the count rates have been corrected
for the increasing deadtime that occurs at very high fluxes. For reference, the
beginning of the observation corresponds to MJD 57197.935. After the first four
\nustar\ orbits, extreme flaring is observed with incident count rates exceeding
10,000\,\ctps\ on several occasions. The strongest six flares, analysed in Section
\ref{sec_flares}, are highlighted (red numbers). We also show the evolution of a
broadband hardness ratio, computed between 3--10 and 10--79\,keV (bottom panel).
Strong spectral variability is observed throughout this latter flaring phase.
}
\vspace{0.3cm}
\label{fig_lcHR}
\end{figure*}

\section{Observations and Data Reduction}
\label{sec_red}

Triggered by the summer 2015 outburst, we undertook five observations with \nustar.
The timing of these observations is shown in the context of the long-term variability
seen by \integral\ in Figure \ref{fig_longlc}; the first was undertaken during the height
of the activity from the source, and the remaining four were spaced throughout the
following few weeks (\citealt{Walton15fade2, Walton15fade1}), during which \v404\ 
declined back to quiescence (\citealt{Sivakoff15quies, Sivakoff15fade}). In this work,
we focus on the first observation. Although this is split over two OBSIDs
(90102007002, 90102007003), in reality they comprise one continuous observation.
The subsequent \nustar\ observations will be presented in Rana et al. (in preparation).

The \nustar\ data were reduced largely following standard procedures. Unfiltered
event files were cleaned using \nupipeline, part of the \nustar\ Data Analysis Software
(v1.5.1; part of the standard HEASOFT distribution), and instrumental responses from
\nustar\ CALDB v20150316 are used throughout this work. Due to the high count rate
and rapid variability, it was necessary to turn off some of the filtering for hot pixels
normally performed by \nupipeline, since source counts were being removed from the
peak flares. We did this by setting the `statusexpr' parameter to
``b0000xx00xx0xx000'', which controls the filtering on the STATUS column. In this
way we kept the source events that were incorrectly identified as hot/flickering. The
\nustar\ calibration database has a list of hot/flickering pixels that have already been
identified, which were still removed following standard procedures. Passages of
\nustar\ through the South Atlantic Anomaly were also excluded from our analysis.

Source products were then extracted from the cleaned events from a circular region
centered on the source (radius 160$''$) using \nuproducts\ for both focal plane
modules (FPMA and FPMB). \v404\ is easily detected across the whole 3--79\,\kev\
\nustar\ bandpass. Owing to its extreme brightness, there were no regions of the
detector on which \v404\ was located that were free of source counts, so the
background was estimated from a blank region on the detector furthest from the
source position (each FPM contains four detectors in a $2 \times 2$ array) in order to
minimize any contribution from the source to our background estimation. Although
there are known to be variations in the background between the detectors for each
FPM, these differences are typically only at the ~10\% level (in the background rate)
at the highest energies of the \nustar\ bandpass (where the internal detector
background dominates; \citealt{NUSKYBKG}). \v404\ is always a factor of $>$10
above the estimated background at all energies in the spectra extracted here, so
such effects are negligible. Finally, when necessary, data from the two OBSIDs
were combined using \addascaspec\ for each FPM (although we do not combine
the FPMA and FPMB data), and all spectra were grouped such that each spectral
bin contains at least 50 counts per energy bin, to allow the use of \chisq\
minimization during spectral fitting.

\section{Temporal Variability}
\label{sec_time}

In Figure \ref{fig_lcHR} (\textit{top panel}), we show the lightcurve observed by
\nustar. The count rate shown is the incident count rate inferred rather than that
directly recorded, i.e. the rate has been corrected for the deadtime (see
\citealt{NUSTAR, Bachetti15}). The most striking aspect is the strong flaring seen
throughout the majority of the observation, during which the flux observed from
\v404\ can rapidly increase by at least an order of magnitude. Many flares
comfortably exceed rates of 10,000 \ctps\ (unless stated otherwise, count rates
are quoted per FPM), with the most extreme even exceeding 20,000 \ctps. For
reference, the incident 3--79\,keV count rate for the Crab nebula is $\sim$500
\ctps\ (\citealt{Madsen15}). Strong X-ray flaring from \v404\ has been reported by
several authors throughout this outburst (\eg\ \citealt{Rodriguez15v404, Natalucci15,
Roques15, King15v404, Jenke16v404, SanFan16}). We stress again that even at
these count rates the \nustar\ data do not suffer significantly from pile-up; at similar
count rates Sco X-1 only had a pile-up fraction of $\sim$0.08\% (see appendix C in
\citealt{Gref16solar}).

In addition to the extreme flux variability, we also see strong spectral variability
throughout the \nustar\ observation. Figure \ref{fig_lcHR} (\textit{bottom panel})
shows the evolution of a simple broadband hardness ratio, computed as the ratio
between the count rates in the  3--10 and 10--79\,keV energy bands, which shows
a remarkable transition between the 4th and 5th \nustar\ orbits. During the first four
orbits, the hardness ratio is relatively stable, but after this point it becomes strongly
variable. This transition is roughly coincident with the onset of the flaring portion of
the observation. The data from the first four orbits will be discussed in more detail
in a dedicated paper (Walton et al. in preparation); here we focus on the strong
flaring seen throughout the majority of the \nustar\ observation.

\begin{figure}
\hspace*{-0.5cm}
\epsscale{1.15}
\plotone{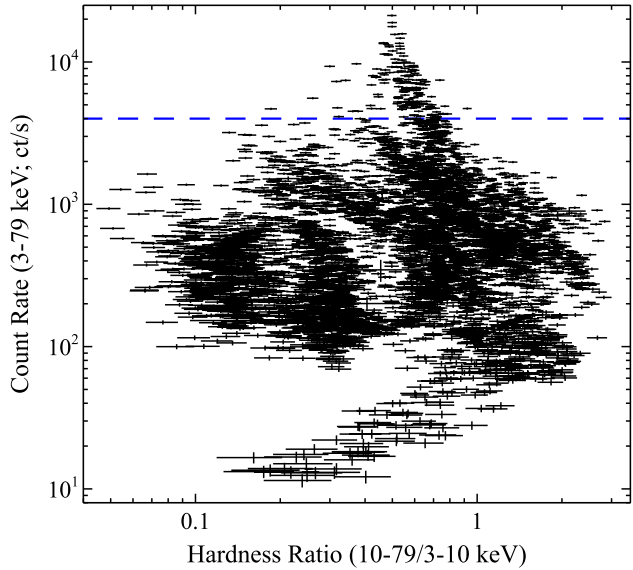}
\caption{Hardness ratio--intensity diagram constructed from the data shown in
Figure \ref{fig_lcHR}. The behaviour seen during this \nustar\ observation is
extremely complex. However, the strongest flares all show similar hardness ratios.
The dashed blue line marks the count rate limit adopted in extracting the flare
spectra discussed in Sections \ref{sec_avflares} and \ref{sec_flares}.
}
\vspace{0.3cm}
\label{fig_hri}
\end{figure}

In order to further characterise the observed variability, in Figure \ref{fig_hri} we plot
the 10--79/3--10\,\kev\ hardness ratio against the full 3--79\,\kev\ count rate. The
resulting `hardness ratio -- intensity' (HRI) diagram is rather chaotic, with no clear
single trend and a lot of complex structure. There are two distinct `clouds' at moderate
intensity with softer spectra (lower hardness ratio, $\lesssim$0.4), which primarily
correspond to the data from the first four \nustar\ orbits. The more complex behaviour
seen in the rest of the data arises from the flaring period. Noteably, though, the flares
themselves all appear to have similar hardness ratios. Finally, at the very lowest
fluxes observed there also appears to be a clear positive correlation between flux and
hardness ratio, which breaks down above $\sim$100\,\ctps.

\section{Spectral Analysis}
\label{sec_spec}

The majority of this work focuses on spectral analysis of data extracted from the period
of intense flaring observed by \nustar. Our spectral analysis is performed with \xspecv\
(\citealt{xspec}), and parameter uncertainties are quoted at 90\% confidence for one
parameter of interest throughout this work (\ie $\Delta\chi^{2} = 2.71$). Residual cross
calibration flux uncertainties between the FPMA and FPMB detectors are accounted for
by allowing multiplicative constants to float between them, fixing FPMA to unity; the
FPMB constants are always found to be within 5\% of unity, as expected
(\citealt{NUSTARcal}).

\begin{figure}
\hspace*{-0.5cm}
\epsscale{1.15}
\plotone{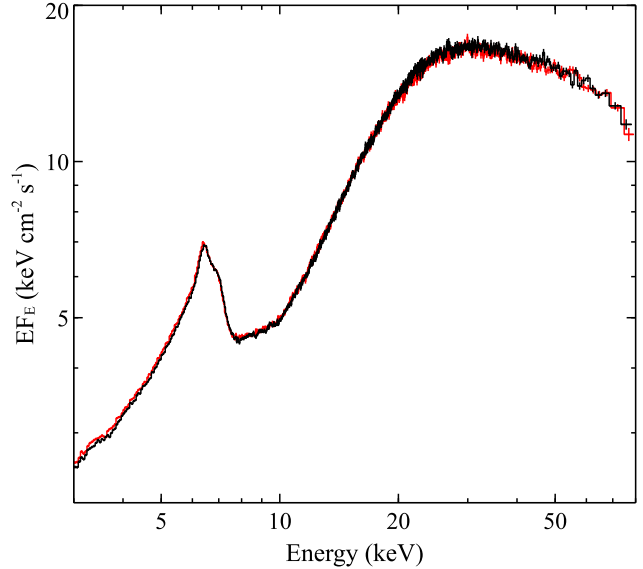}
\caption{The average X-ray spectrum from our first \nustar\ observation of \v404.
FPMA data are shown in black, and FPMB data in red; both have been unfolded
through a model that is constant with energy, and have been further rebinned for
visual purposes. While strong spectral variability is observed throughout the
observation, the average spectrum is still useful for highlighting certain features,
noteably a narrow iron emission component, indicating the presence of
reprocessing by distant material, and a strong absorption edge at $\sim$7\,keV,
indiciating the presence of absorption significantly in excess of the Galactic
column throughout much of the observation.}
\vspace{0.3cm}
\label{fig_spec_av}
\end{figure}

In Figure \ref{fig_spec_av} we show the average spectrum obtained from the full
\nustar\ observation. Given the strong spectral variability discussed previously, a
detailed analysis of this average spectrum would not be particularly meaningful.
However, a visual inspection is still useful in terms of highlighting some of the
features of the observed data. In particular, there is clear structure in the iron K
bandpass. There is a strong absorption edge above 7\,\kev, indicating there
is absorption in excess of the Galactic column ($N_{\rm{H,Gal}} \sim 10^{22}$
cm$^{-2}$; \eg\ \citealt{Reynolds14quies, Bernardini14, Rana16}) throughout much
of the observation. This is similar to the 1989 outburst (\eg\ \citealt{Oosterbroek97,
Zycki99b}). In addition, as discussed by \cite{King15v404} and \cite{Motta16v404},
there is a clear, narrow emission line from neutral iron, indicating a contribution
from reprocessing by distant, neutral material; evidence for such emission was
also seen in the 1989 data (\citealt{Zycki99b}).

\subsection{The Average Flare Spectrum}
\label{sec_avflares}

\begin{figure}
\hspace*{-0.5cm}
\epsscale{1.15}
\plotone{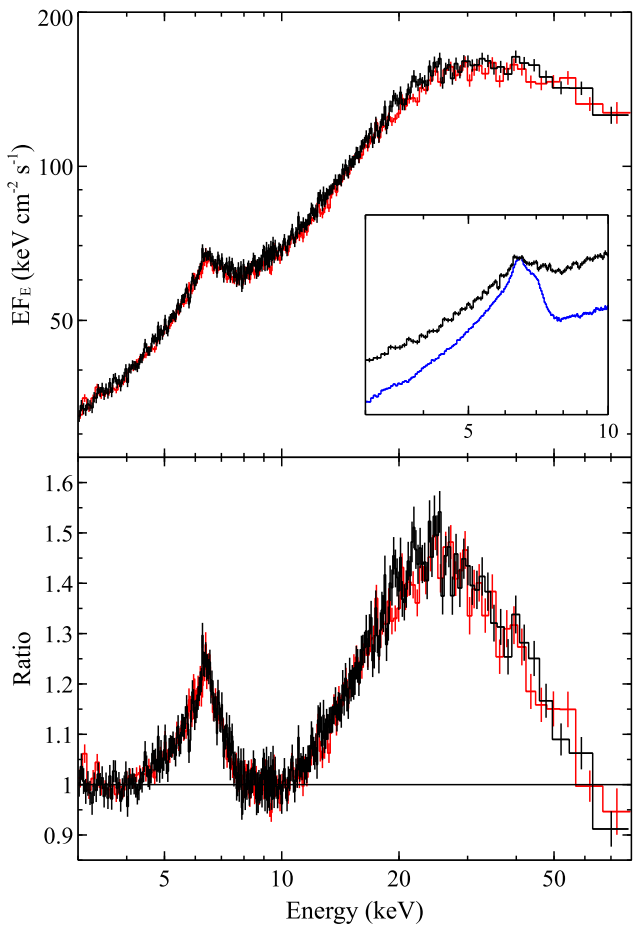}
\caption{The flare spectrum, extracted from periods where the count rate (per FPM)
exceeds 4,000\,\ctps\ (top panel, computed in the same manner as Figure
\ref{fig_spec_av}). As before, the FPMA and FPMB data are shown in black and
red, respectively, and the data have been further rebinned for visual purposes. The
inset shows a comparison between the FPMA data for the flare spectrum and the
average spectrum (blue) in the iron K bandpass, with the latter scaled up in flux so
that the peaks of the narrow iron emission match; the strong edge seen in the
average spectrum is not present in the flare spectrum. The bottom panel shows the
data/model ratio to a simple powerlaw continuum with a high-energy exponential
cutoff, fit to the 3--4, 8--10 and 50--79\,keV bands. The residuals imply the
presence of a strong reflection component from the inner accretion disk.}
\vspace{0.3cm}
\label{fig_flares}
\end{figure}

In Figure \ref{fig_flares} (\textit{top panel}), we show the average spectrum for the
flares, extracted by selecting only periods where the count rate (per FPM) was $>$
4000\,\ctps. The total good exposure in the resulting spectrum is only
$\sim$110--120\,s. In contrast to the average spectrum, there is no visually apparent
edge at $\sim$7\,\kev, indicating the line-of-sight absorption is much weaker during
these periods, and that we therefore have a cleaner view of the intrinsic spectrum.
The flare spectrum is very hard, and there is still visible structure in the iron K band.
In Figure \ref{fig_flares} (\textit{bottom panel}), we show the data/model residuals to
a simple model consisting of a powerlaw continuum with a high-energy exponential
cutoff, modified by a neutral absorption column which is free to vary above a lower
limit of $10^{22}$\,cm$^{-2}$ (set by prior constraints on the Galactic column; see
above). We use the \tbabs\ absorption model, adopting the ISM abundances
reported in \cite{tbabs} as our `solar' abundance set, and the cross-sections of
\cite{Verner96}, as recommended. This model is fit to the 3--4, 8--10 and 50--79\,keV
energy ranges in order to minimize the influence of any reflected emission present in
the spectrum. The photon index obtained is very hard, $\Gamma \sim 1.5$, with a
cutoff energy of $E_{\rm{cut}} \sim 160$\,keV.

%\begin{figure*}
%\hspace*{-0.55cm}
%\epsscale{0.525}
%\plotone{./figs/v404cyg_flares_eeuf_inset.eps}
%\hspace*{1.1cm}
%\plotone{./figs/v404cyg_flares_ratio.eps}
%\caption{
%\nustar\ flare spectrum. Reflectiony.
%}
%\label{fig_flares}
%\end{figure*}

A very strong Compton hump is visible around $\sim$20--30\,keV, indicating a
significant contribution from X-ray reprocessing by optically-thick material. The
iron emission is also rather strong, and although there is a narrow core to the line
profile, the majority of the line emission is broadened with a clear red-wing, a
hallmark of relativistically broadened reflection from an accretion disk (referred to
as a `diskline' profile; \eg\ \citealt{Fabian89, kdblur}). Modeling the 3--10\,\kev\
bandpass with the simple continuum model above (fixing the cutoff energy to its
best-fit value, given the limited energy range being considered), and including
both an unresolved Gaussian at 6.4\,keV and a \relline\ component
(\citealt{relconv}) to account for the narrow core and the iron emission from the
accretion disk, respectively, we find that the \relline\ component has an equivalent
width of $\mathrm{EW} \sim 400$\,eV, while the narrow core is much weaker, with
$\mathrm{EW} \sim 25$\,eV.

\begin{figure}
\hspace*{-0.5cm}
\epsscale{1.15}
\plotone{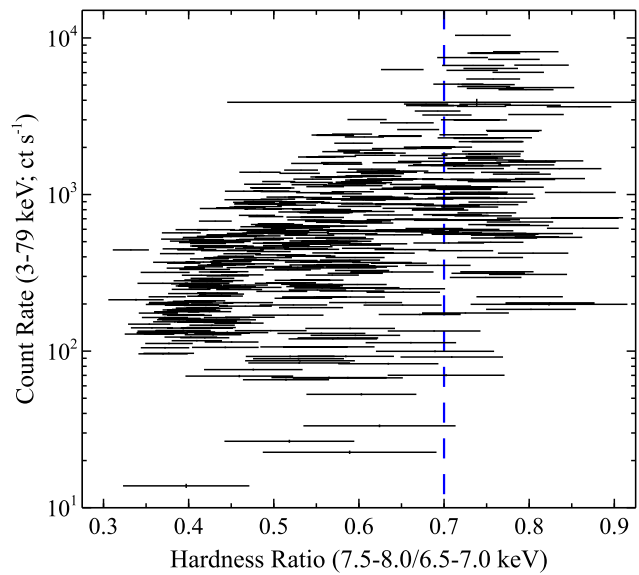}
\caption{Hardness ratio--intensity diagram, similar to Figure \ref{fig_hri} but with
100s time bins, for the narrow-band hardness ratio (\hredge; see Section
\ref{sec_sel}) constructed to probe the depth of the iron edge at $\sim$7\,keV. The
major flares, which do not show the prominent edge seen in the average spectrum
(Figures \ref{fig_spec_av}, \ref{fig_flares}), show \hredge\ $>$ 0.7 (indicated with
the dashed blue line), which is used as a limit to identify other periods with similarly
low levels of absorption.
}
\vspace{0.3cm}
\label{fig_hri_edge}
\end{figure}

\subsection{Flux-Resolved Spectral Evolution}
\label{sec_flux}

Isolating and modeling the reprocessed emission from the accretion disk is of
significant importance, as this provides information on both the spin of the black
hole (\eg\ \citealt{Risaliti13nat, Walton13spin, Walton14, Reynolds14rev}), and the
geometry/location of the illuminating X-ray source (\eg\ \citealt{Wilkins12}). This is
of particular interest for the intense flares, since such X-ray flares are often
associated with jet ejection (\eg\ \citealt{Corbel02}). However, constraining the disk
reflection is not necessarily straightforward from the iron band alone. In order to
aid in disentangling the contributions from reprocessing by the accretion disk and
by more distant material to the  spectrum, the main body of this work
focuses on modeling the broadband evolution of \v404\ as a function of flux during
the flaring phase of our \nustar\ observation.

\subsubsection{Data Selection}
\label{sec_sel}

One of the main complications for broadband modeling is the strong and variable
absorption that is present throughout this observation. In order to mimimize this
issue, based on the flare spectrum (Figure \ref{fig_flares}), we select only
periods of similarly low absorption for our flux-resolved spectral analysis. In order to
identify such periods, we define a narrow band hardness ratio (hereafter \hredge),
with the softer band (6.5--7.0\,\kev) just below the sharp edge seen in the average
spectrum, and the harder band (7.5--8.0\,\kev) just above, such that we can track
the strength of the edge throughout the flaring period. With these narrow bands, a
stronger absorption edge (and thus more absorption) would appear to have a
softer spectrum (i.e. a lower hardness ratio). We show the behaviour of \hredge\ in
Figure \ref{fig_hri_edge}, in the form of a similar HRI diagram to Figure \ref{fig_hri}.
Note that we are forced to adopt a coarser temporal binning (100s) in order for
\hredge\ to be well constrained owing to the narrow energy bands used; hence the
peak 3--79\,keV count rates differ in this Figure. Nevertheless, it is clear that in
terms of \hredge, the highest count rates (\ie\ the strongest flares) show the hardest
spectra, with \hredge\ $\gtrsim 0.7$, consistent with the lack of absorption seen in
Figure \ref{fig_flares}. Furthermore, although the majority of the observation shows
a much stronger edge, there are other non-flare periods in which the absorption is
similarly weak. These periods are spread randomly throughout the flaring portion of
the observation, and span a broad range of flux.

\begin{figure}
\hspace*{-0.5cm}
\epsscale{1.15}
\plotone{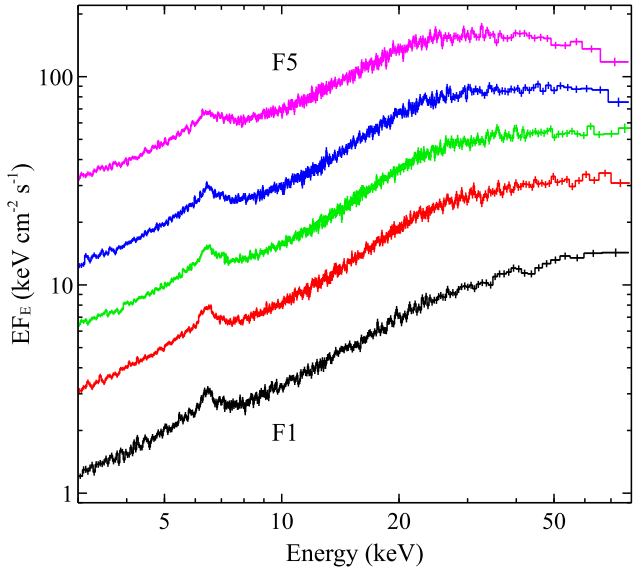}
\caption{The five X-ray spectra extracted from periods of low absorption
(determined based on the strength of the absorption edge at $\sim$7\,keV) for our
flux-resolved analysis (F1--5, shown in black, red, green, blue and magenta,
respectively; the highest flux state, F5, is the same as the flare spectrum shown in
Figure \ref{fig_flares}). Only the FPMA data are shown for clarity, and as with Figures
\ref{fig_spec_av} and \ref{fig_flares}, the data have been unfolded through a 
constant, and the data have been rebinned for visual purposes. The X-ray spectrum
visibly evolves with flux, with the continuum above $\sim$10\,\kev\ becoming more
peaked (i.e. there is more spectral curvature) at higher fluxes.
}
\vspace{0.3cm}
\label{fig_5lvl}
\end{figure}

We therefore select only data with \hredge\ $\geq 0.7$ for the lower fluxes intervals
(i.e. $<$4000\,\ctps), and then divide these periods into four flux bins: 100--500,
500--1000, 1000--2000 and 2000--4000\ \ctps\ (per FPM, using the count rates
from the finer 10\,s binning). The lower limit to the data considered is set to
100\,\ctps\ in order to avoid the low flux region in which the flux and the broadband
hardness ratio are correlated (see Figure \ref{fig_hri}), as the source behaviour is
clearly distinct in this regime. We therefore have five flux bins in total (referred to as
F1--5, in order of increasing flux), including the flare spectrum extracted from
$>$4000\,\ctps\ in section \ref{sec_avflares}. Details of these flux bins are given in
Table \ref{tab_flux} and the extracted spectra are shown in Figure \ref{fig_5lvl};
the lack of strong, visible absorption edges in any of these spectra demonstrates
the general success of our low-absorption selection procedure. As with the flare
spectrum, despite the lack of strong absorption, the spectra from lower fluxes are
also very hard. There are a couple of trends that can be seen from a visual
inspection of these data. First, the relative contribution from the narrow core of the
iron emission is stronger at lower fluxes. Second, the continuum above
$\sim$10\,\kev\ is shows a lot more spectral curvature at higher fluxes.

Before proceeding with our more detailed spectral analysis, we repeat our
phenomenological modelling of the 3--10\,\kev\ bandpass performed above for the
flare spectrum, and fit the data for each of these flux bins with a combination of a
broad and narrow iron emission component. In order to minimize parameter
degeneracies, given the limited bandpass utilized, we make the simplifying
assumption that the profile of the broad iron emission is the same for all fluxes.
With this simple modelling, although the individual uncertainties are relatively large,
we find that the strength of the broad iron emission increases with increasing flux
(see Table \ref{tab_flux}).

\subsubsection{Basic Model Setup}
\label{sec_mod}

Having extracted our low-absorption, flux-resolved \nustar\ data, we construct a
spectral model for \v404\ that incorporates both the primary emission from the
black hole, as well as X-ray reprocessing by both the accretion disk, and more
distant material. Our model also includes neutral absorption, allowing for both the
Galactic column and a second absorption column, assumed to be intrinsic to the
source, to account for any absorption in excess of the Galactic column.

To model the relativistic disk reflection, we use the \relxill\ model (\citealt{relxill}).
This is a merging of the \xillver\ reflection model (v0.4c; \citealt{xillver}) with the
\relconv\ model for the relativistic effects close to a black hole that smear out the
rest-frame reflection spectrum (\citealt{relconv}). In particular, given the potential
association between the X-ray flares and jet activity (as mentioned above, and
discussed in more detail in section \ref{sec_jet}), we use the \relxilllp\ model
(part of the broader \relxill\ family of models). This includes both the primary
continuum --- assumed to be a powerlaw with a high-energy exponential cutoff ---
and the reflected emission from the accretion disk, and treats the illuminating
X-ray source as a point source located above the accretion disk on the spin-axis
of the black hole (\ie\ the `lamppost' accretion geometry), an idealized geometrical
approximation appropriate for the scenario in which the hard X-ray continuum is
associated with the base of a jet (\eg\ \citealt{Markoff05, Miller12}).

The key parameters for the \relxilllp\ model are the photon index and the
high-energy cutoff of the illuminating continuum ($\Gamma$, $E_{\rm{cut}}$), the
spin of the black hole ($a^*$), the inclination and the inner and outer radii of the
accretion disk ($i$, \rin, \rout), the iron abundance and ionization parameter of
the accreting material ($A_{\rm{Fe}}$, $\xi = 4\pi F/n$, where $F$ is the ionizing
flux incident on the disk, integrated between 1--1000\,Ry, and $n$ is the density of
the material), the height of the illuminating source above the disk ($h$) and the
strength of the disk reflection ($R_{\rm{disk}}$). Note that here, we use the
``reflection fraction" definition outlined in \cite{relxill_norm}. This determines the
strength of the reflected emission from the relative intensities of the powerlaw
continuum as seen by the disk and by the distant observer, which can be 
computed self-consistently for the lamppost geometry via relativistic ray-tracing.
The outer radius of the disk is set to 1000\,\rg\ throughout our analysis (where
\rg\ is the gravitational radius), the maximum permitted by the model, and
following \cite{Garcia15}, we consider cutoff energies up to 1000\,keV. We also
compute $h$ in units of the event horizon (\rh, which varies between 1 and 2\,\rg\ 
for maximally rotating and non-rotating black holes, respectively) throughout this
work, so that we can require that the X-ray source is always outside this radius.
For practical reasons, we actually set a lower limit of 2\rh\ for $h$ in order to
prevent the model from implying unphysically small X-ray sources, as the
illuminating source obviously must have some physical extent (particularly if it is
associated with a jet) despite being approximated in our models as a point source.

\begin{table}
  \caption{Details of the five flux bins used in our flux resolved spectral analysis of
  the peruids of low absorption (selected to have \hredge\ $\geq$ 0.7).}
  \vspace{-0.25cm}
\begin{center}
\begin{tabular}{c c c c c}
\hline
\hline
\\[-0.1cm]
Flux & Count & Good & Broad \\
\\[-0.25cm]
Bin & Rate & Exposure & Fe K EW \\
\\[-0.25cm]
& (ct s$^{-1}$ FPM$^{-1}$) & (FPMA/B; s) & (eV) \\
\\[-0.2cm]
\hline
\hline
\\[-0.1cm]
F1 & 100--500 & 1074/1105 & $260^{+70}_{-60}$ \\
\\[-0.2cm]
F2 & 500--1000 & 1067/1120 & $350^{+60}_{-50}$ \\
\\[-0.2cm]
F3 & 1000-2000 & 722/769 & $390^{+40}_{-80}$ \\
\\[-0.2cm]
F4 & 2000-4000 & 260/280 & $460^{+60}_{-100}$ \\
\\[-0.2cm]
F5 & $>$4000 & 112/121 & $440^{+50}_{-90}$ \\
\\[-0.2cm]
\hline
\hline
\end{tabular}
\vspace{-0.2cm}
\label{tab_flux}
\end{center}
\vspace{0.5cm}
\end{table}

For the distant reprocessor, we use the \xillver\ reflection model. As the narrow
core is at 6.4\,keV, we assume this to be neutral (i.e. $\log\xi = 0$; throughout this
work we quote $\xi$ in units of \ergcmps). The key parameters here are the photon
index and high-energy cutoff of the illuminating continuum, the inclination of the
reflecting slab, the iron abundance, and the strength of the reflected emission. Both
the photon index and the high-energy cutoff are assumed to be the same as for the
\relxilllp\ component, and as \relxilllp\ already includes the primary continuum
emission, we configure the \xillver\ model to only provide the reflected emission (\ie\ 
we set the reflection fraction parameter to $-1$). One complication is that the
geometry of the distant reprocessor is not known, and different geometries can
result in differences in the reflected spectra (\eg\ \citealt{Brightman15}). \xillver\ 
assumes a simple semi-infinite slab, but this is unlikely to be physically realistic.
Therefore in order to allow the \xillver\ component representing the distant
reprocessor the flexibility to differ from the simple slab approximation, we allow the
iron abundance and inclination parameters of this component to vary independently
of the other model components. These are effectively `dummy' parameters which
allow us to incorporate this flexibility with a simple parameterization. However, we
set a lower limit on $A_{\rm{Fe}}$ of \felim, such that the limit in which the distant
reflection dominates the 2--10\,keV bandpass would remain consistent with
\cite{King15v404}, who report equivalent widths of up to 1\,keV for the narrow,
neutral iron emission based on their analysis of the high-resolution \chandra\
HETG data taken during this outburst.

\begin{table*}
  \caption{A summary of the lamppost reflection models applied during our
  flux- and flare-resolved analyses, presented in Sections \ref{sec_flux} and
  \ref{sec_flares}, respectively.}
  \vspace{-0.25cm}
\begin{center}
\begin{tabular}{c c c c}
\hline
\hline
\\[-0.1cm]
Model & Dataset & Source Emission & Notes \\
\\[-0.2cm]
\hline
\\[-0.1cm]
1 & Flux-resolved & Lamppost only & $R_{\rm{disk}}$ a free parameter, \rin\ fixed at the ISCO \\
\\[-0.2cm]
2 & Flux-resolved & Lamppost only & $R_{\rm{disk}}$ calculated self-consistently, \rin\ free to vary, $h$ constant \\
\\[-0.2cm]
3 & Flux-resolved & Lamppost only & $R_{\rm{disk}}$ calculated self-consistently, \rin\ fixed at the ISCO, $h$ free to vary \\
\\[-0.2cm]
4 & Flux-resolved & Lamppost $+$ disk & $R_{\rm{disk}}$ calculated self-consistently, \rin\ free to vary, $h$ constant \\
\\[-0.2cm]
5 & Flare-resolved & Lamppost only & $R_{\rm{disk}}$ calculated self-consistently, \rin\ fixed at the ISCO \\
\\[-0.2cm]
6 & Flare-resolved & Lamppost $+$ disk & $R_{\rm{disk}}$ calculated self-consistently, \rin\ fixed at the ISCO \\
\\[-0.2cm]
6i & Flare-resolved & Lamppost $+$ disk & Same as model 6, but $i$ limited to $\geq$50\deg\ \\
\\[-0.2cm]
\hline
\hline
\end{tabular}
\vspace{-0.2cm}
\label{tab_model}
\end{center}
\end{table*}

\begin{table*}
  \caption{Results for the free parameters in the basic lamppost reflection
  model (Model 1) constructed for the spectral evolution as a function of flux.}
  \vspace{-0.25cm}
\begin{center}
\begin{tabular}{c c c c c c c c c}
\hline
\hline
\\[-0.1cm]
Model Component & \multicolumn{2}{c}{Parameter} & Global & \multicolumn{5}{c}{Flux Level} \\
\\[-0.15cm]
& & & & F1 & F2 & F3 & F4 & F5 \\
\\[-0.2cm]
\hline
\hline
\\
\tbabs$_{\rm{src}}$ & \nh\ & [$10^{21}$ cm$^{-2}$] & & $9.1^{+2.2}_{-1.7}$ & $9.1^{+1.6}_{-2.1}$ & $<0.7$ & $<1.6$ & $<0.4$ \\
\\[-0.2cm]
\relxilllp\ & $\Gamma$ & & & $1.43 \pm 0.02$ & $1.49 \pm 0.03$ & $1.42^{+0.02}_{-0.01}$ & $1.41^{+0.02}_{-0.01}$ & $1.40^{+0.01}_{-0.02}$ \\
\\[-0.2cm]
& $E_{\rm{cut}}$ & [keV] & & $>840$\tmark[a] & $610^{+340}_{-210}$ & $240^{+10}_{-20}$ & $150^{+20}_{-10}$ & $120\pm10$ \\
\\[-0.2cm]
& $a^*$ & & $>-0.1$ \\
\\[-0.2cm]
& $i$ & [\deg] & $27\pm2$ \\
\\[-0.2cm]
& $h$ & \rh\ & & $6.0^{+7.0}_{-2.0}$ & $4.7^{+4.0}_{-1.0}$ & $3.9^{+3.0}_{-1.1}$ & $3.7^{+2.8}_{-0.9}$ & $3.2^{+2.5}_{-1.1}$ \\
\\[-0.2cm]
& $\log\xi$ & $\log$[\ergcmps] & & $3.01^{+0.02}_{-0.01}$ & $3.02 \pm 0.01$ & $3.09^{+0.01}_{-0.02}$ & $3.15^{+0.05}_{-0.02}$ & $3.47^{+0.05}_{-0.04}$ \\
\\[-0.2cm]
& $A_{\rm{Fe}}$ & [solar] & $1.9^{+0.3}_{-0.1}$ \\
\\[-0.2cm]
& $R_{\rm{disk}}$ & & & $1.1\pm0.2$ & $1.5^{+0.3}_{-0.2}$ & $1.7^{+0.4}_{-0.2}$ & $2.0^{+0.6}_{-0.2}$ & $3.0^{+0.8}_{-0.5}$ \\
\\[-0.2cm]
& Norm & & & $0.15^{+0.03}_{-0.02}$ & $0.23^{+0.05}_{-0.03}$ & $0.33^{+0.08}_{-0.09}$ & $0.47^{+0.09}_{-0.06}$ & $0.65^{+0.34}_{-0.15}$ \\
\\[-0.2cm]
\xstar$_{\rm{abs}}$ & $\log\xi$ & $\log$[\ergcmps] & $4.6^{+0.8}_{-0.3}$ \\
\\[-0.2cm]
& \nh\ & [$10^{21}$ cm$^{-2}$] & & $3.7^{+5.0}_{-2.3}$ & $4.2^{+4.8}_{-1.5}$ & $<3.5$ & $3.4^{+4.0}_{-1.3}$ & $3.0^{+2.7}_{-1.1}$ \\
\\[-0.2cm]
\xillver\ & $i$\tmark[b] & [\deg] & $<11$ \\
\\[-0.2cm]
& $A_{\rm{Fe}}$\tmark[b] & [solar] & $<0.91$ \\
\\[-0.2cm]
& Norm & [$10^{-2}$] & & $1.2^{+1.1}_{-0.8}$ & $8.6^{+0.7}_{-1.1}$ & $12.7^{+0.8}_{-0.7}$ & $18.7^{+1.6}_{-1.5}$ & $33.0^{+2.7}_{-3.0}$ \\
\\[-0.2cm]
\xstar$_{\rm{emis}}$ & $\log\xi$ & $\log$[\ergcmps] & & $<1.7$ \\
\\[-0.2cm]
& Norm & [$10^{4}$] & & $1.3\pm0.3$ \\
\\[-0.2cm]
\hline
\\[-0.1cm]
\chisq/DoF & & & 10599/10308 & \\
\\[-0.2cm]
\hline
\\[-0.15cm]
$F_{3-79}$\tmark[c] & \multicolumn{2}{c}{[$10^{-8}$\,\ergpcmsqps]} & & $3.78 \pm 0.02$ & $8.85 \pm 0.03$ & $16.06 \pm 0.06$ & $28.0 \pm 0.1$ & $54.6 \pm 0.3$ \\
\\[-0.2cm]
\hline
\hline
\end{tabular}
\vspace{-0.2cm}
\label{tab_param}
\end{center}
\vspace{0.1cm}
$^a$ $E_{\rm{cut}}$ is constrained to be $\leq$1000\,keV following \cite{Garcia15}. \\
$^b$ These act as dummy `shape' parameters to allow this component the
flexibility to deviate from the simple slab approximation adopted in the XILLVER
model (see main text), and in turn allow for our lack of knowledge with regards to
the geometry of the distant reprocessor. \\
$^c$ Average flux in the 3--79\,keV bandpass
\vspace{0.4cm}
\end{table*}

\cite{King15v404} also find both emission and absorption lines from \fexxv\ and
\fexxvi. We therefore also allow for a contribution from photoionized emission and
absorption. These are treated with grid models generated with \xstar\
(\citealt{xstar}), and are customised specifically for \v404. In brief, these grids are
calculated assuming the abundances set derived by \cite{GonzHern11} for the
stellar companion of \v404, and their free parameters are the ionization state of
the material, its column density, and its outflow velocity (see King et al., in
preparation, for full details). While \cite{King15v404} find the absoption features to
be mildly outflowing, the velocity shifts are small in comparison to the spectral
resolution of \nustar, and we therefore keep these photoionised components to be
fixed at rest.

\begin{figure}
\hspace*{-0.5cm}
\epsscale{1.15}
\plotone{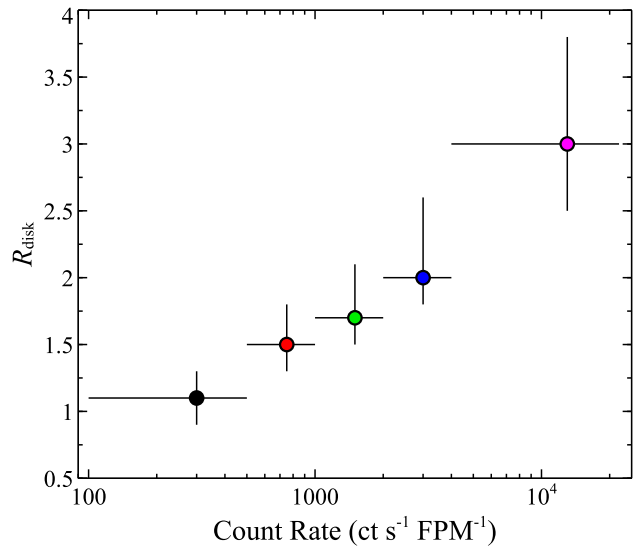}
\caption{The evolution of the disk reflection fraction $R_{\rm{disk}}$ with source
flux inferred from our basic lamppost model for the spectral evolution seen from
\v404\ (Model 1). We find that $R_{\rm{disk}}$ increases with increasing flux,
implying an evolution in the geometry of the innermost accretion flow. The strong
reflection found at the highest fluxes ($R_{\rm{disk}} \sim 3$) would require
gravitational lightbending. The data points are color-coded to match the spectra
shown in Figure \ref{fig_5lvl}.
}
\label{fig_Rfrac}
\end{figure}

Finally, for the neutral absorption, we again use the \tbabs\ model. The Galactic
column is set to $N_{\rm{H,Gal}} = 10^{22}$\,cm$^{-2}$, as discussed previously,
and is assumed to have the ISM abundances of \cite{tbabs}. For the additional,
source intrinsic absorption, we use the version of \tbabs\ with variable elemental
abundances, so that we can link the iron abundance of this absorber to that of
the disk reflection model (\ie\ we assume that \v404\ is a chemically homogeneous
system). We assume that this absorber is sufficiently distant from the innermost
accretion flow that it should act on all of the emission components arising from this
region. This absorber may potentially be associated with the distant reprocessor,
and to allow for this possibility we configure the model such that while the Galactic
absorption acts on all the emission components, the source intrinsic absorber acts
only on the primary emission and the relativistic disk reflection, but not the distant
reflection. However, this choice with makes little difference to the results obtained,
as the distant reprocessor makes a negligible contribution to the spectrum at the
lowest energies covered by \nustar. The form of the basic model applied, in \xspec\
jargon, is therefore as follows: \tbabs$_{\rm{Gal}}$ $\times$ ( \xillver\ $+$
\xstar$_{\rm{emis}}$ $+$ (  \xstar$_{\rm{abs}}$ $\times$ \tbabs$_{\rm{src}}$
$\times$ \relxilllp\ ) )

We apply this model to the five flux states shown in Figure \ref{fig_5lvl}
simultaneously. In doing so, we require the black hole spin, the inclination of the
accretion disk and the iron abundance of the system to be the same across all
flux states, as these physical parameters should not vary over the course of this
\nustar\ observation. As it is unlikely that the geometry of the distant reprocessor
would evolve significantly throughout our observation, we also link the `shape'
parameters that would relate to geometry in our simple parameterisation (iron
abundance, slab inclination) for this component across all flux levels. However,
we do allow this component to respond to the changes in the intrinsic emission 
from \v404. During the fitting process we found that the ionizaion of the \xstar\
absorption component was consistent for all the flux states, so for simplicity we
also linked this parameter. Additionally, we found that the photoionized emission
only makes a significant contribution to the lowest flux state, F1, and so fixed its
normalization to zero for F2--5. Furthermore, we found that this photoionised
emission only provided an additional contribution to the narrow Fe K emission,
and as such the column density and normalisation were highly degenerate, so
we fixed the former to an arbitrary value of $10^{19}$\,cm$^{-2}$.

\begin{table*}
  \caption{Results for the high-spin solutions obtained with the lamppost
  reflection models constructed to investigate potential geometric evolution scenarios
  as a function of flux (Models 2 and 3).}
  \vspace{-0.25cm}
\begin{center}
\begin{tabular}{c c c c c c c c c}
\hline
\hline
\\[-0.1cm]
Model Component & \multicolumn{2}{c}{Parameter} & Global & \multicolumn{5}{c}{Flux Level} \\
\\[-0.15cm]
& & & & F1 & F2 & F3 & F4 & F5 \\
\\[-0.2cm]
\hline
\hline
\\[-0.1cm]
\multicolumn{9}{c}{Model 2: truncating disk, static corona} \\
\\[-0.1cm]
\relxilllp\ & $\Gamma$ & & & $1.41^{+0.02}_{-0.03}$ & $1.44^{+0.02}_{-0.03}$ & $1.40\pm0.01$ & $1.37\pm0.02$ & $1.37\pm0.01$ \\
\\[-0.2cm]
& $E_{\rm{cut}}$ & [keV] & & $>540$\tmark[a] & $330\pm60$ & $190\pm10$ & $125^{+8}_{-6}$ & $91^{+4}_{-3}$ \\
\\[-0.2cm]
& $a^*$ & & $>0.95$ \\
\\[-0.2cm]
& $i$ & [\deg] & $36\pm1$ \\
\\[-0.2cm]
& $h$ & \rh\ & $2.3^{+0.4}_{-0.1}$ \\
\\[-0.2cm]
& $A_{\rm{Fe}}$ & [solar] & $3.0\pm0.1$ \\
\\[-0.2cm]
& \rin\ & \risco\ & & $2.5^{+0.4}_{-0.3}$ & $2.3^{+0.1}_{-0.2}$ & $2.0\pm0.1$ & $1.7^{+0.2}_{-0.1}$ & 1 (fixed) \\
\\[-0.2cm]
& $R_{\rm{disk}}$\tmark[b] & & & 1.3 & 1.5 & 1.7 & 1.9 & 3.0 \\
\\[-0.2cm]
& Norm & & & $0.61^{+0.11}_{-0.10}$ & $0.80^{+0.18}_{-0.16}$ & $1.07\pm0.15$ & $1.53^{+0.24}_{-0.31}$ & $2.61^{+0.19}_{-0.51}$ \\
\\[-0.2cm]
\hline
\\[-0.1cm]
\chisq/DoF & & & 10656/10313 & \\
\\[-0.2cm]
\hline
\hline
\\[-0.1cm]
\multicolumn{9}{c}{Model 3: stable disk, dynamic corona} \\
\\[-0.1cm]
\relxilllp\ & $\Gamma$ & & & $1.36^{+0.03}_{-0.01}$ & $1.41 \pm 0.01$ & $1.37^{+0.02}_{-0.01}$ & $1.36 \pm 0.01$ & $1.38 \pm 0.01$ \\
\\[-0.2cm]
& $E_{\rm{cut}}$ & [keV] & & $540^{+80}_{-50}$ & $280 \pm 20$ & $180 \pm 10$ & $123^{+5}_{-3}$ & $94 \pm 3$ \\
\\[-0.2cm]
& $a^*$ & & $>0.88$ \\
\\[-0.2cm]
& $i$ & [\deg] & $28^{+1}_{-2}$ \\
\\[-0.2cm]
& $h$ & \rh\ & & $5.2^{+1.5}_{-0.4}$ & $5.6 \pm 0.3$ & $4.6^{+0.8}_{-0.2}$ & $4.1^{+0.3}_{-0.2}$ & $4.4 \pm 0.2$ \\
\\[-0.2cm]
& $A_{\rm{Fe}}$ & [solar] & $3.03 \pm 0.05$ \\
\\[-0.2cm]
& $R_{\rm{disk}}$\tmark[b] & & & 1.7 & 1.6 & 1.8 & 2.0 & 1.9 \\
\\[-0.2cm]
& Norm & & & $0.18 \pm 0.01$ & $0.22 \pm 0.01$ & $0.38^{+0.05}_{-0.01}$ & $0.63 ^{+0.09}_{-0.04}$ & $1.02^{+0.12}_{-0.07}$ \\
\\[-0.2cm]
\hline
\\[-0.1cm]
\chisq/DoF & & & 10678/10313 & \\
\\[-0.2cm]
\hline
\hline
%\\[-0.1cm]
%\multicolumn{9}{c}{Model 4: stable disk, outflowing corona (varying velocity)} \\
%\\[-0.1cm]
%\relxilllp\ & $\Gamma$ & & &  \\
%\\[-0.2cm]
%& $E_{\rm{cut}}$ & [keV] & &  \\
%\\[-0.2cm]
%& $a^*$ & &  \\
%\\[-0.2cm]
%& $i$ & [\deg] & \\
%\\[-0.2cm]
%& $h$ & \rh\ &  \\
%\\[-0.2cm]
%& $v_{\rm{X}}$ & $c$ & &  \\
%\\[-0.2cm]
%& $A_{\rm{Fe}}$ & [solar] &  \\
%\\[-0.2cm]
%& $R_{\rm{disk}}$\tmark[a] & & &  \\
%\\[-0.2cm]
%& Norm & & &  \\
%\\[-0.2cm]
%\hline
%\\[-0.1cm]
%\chisq/DoF & & & & \\
%\\[-0.2cm]
%\hline
%\hline
\end{tabular}
\vspace{-0.2cm}
\label{tab_param_fix}
\end{center}
$^a$ $E_{\rm{cut}}$ is constrained to be $\leq$1000\,keV following \cite{Garcia15}. \\
$^b$ For these models, $R_{\rm{disk}}$ is calculated self-consistently in the
lamppost geometry from $a^*$, $h$ and \rin. As it is not a free
parameter, errors are not estimated.
\vspace{0.4cm}
\end{table*}

\subsubsection{Results}
\label{sec_res}

To begin with, we assume that the disk extends in to the innermost stable circular
orbit (ISCO) for all flux states and that the corona is not outflowing, and we allow the
reflection fraction to vary as a free parameter (Model 1; a summary of all the models
considered in our flux- and flare-resolved analyses is given in Table \ref{tab_model}).
This model provides a good fit to the global dataset, with \chisq\ = 10599 for 10308
degrees of freedom (DoF). We observe several trends in the fits, which are presented
in Table \ref{tab_param}. Most notably, we find that the strength of the disk reflection 
increases with increasing flux (see Figure \ref{fig_Rfrac}). This is a strong indicator
that the (average) geometry of the innermost accretion flow evolves as a function of
source flux. In addition to these variations, the ionization of the disk increases as the
observed flux increases, as would broadly be expected for an increasing ionizing flux,
and there are changes in the intrinsic continuum, with the high-energy cutoff
decreasing in energy as the flux increases.

The black hole spin is not well constrained with this model (although the majority of
negative spins are excluded: $a^* > -0.1$). However, during the flares the disk 
reflection is very strong, ($R_{\rm{disk}} \sim 3$). Caution over the exact value is
necessary here, as the strength of the reflection obtained is dependent to some
extent on the form of the high-energy curvature included in the input continuum model
(a simple exponential cutoff in this work), and there is also some degeneracy between
the $R_{\rm{disk}}$ and $E_{\rm{cut}}$ parameters. However, taking the result at face
value, this would imply a scenario in which strong gravitational lightbending enhances
the disk reflection (\eg\ \citealt{lightbending}). In turn, this would imply that \v404\
hosts a rapidly rotating black hole (\eg\ \citealt{Parker14mrk, Dauser14}).
Although we are using an idealized lamppost geometry in this work, as long as the
disk is thin then this is the case regardless of the precise geometry of the X-ray source,
as the disk must extend close to the black hole in order to subtend a sufficiently large
solid angle to produce the high reflection fraction; the validity of the thin disk
assumption (which is currently implicit in the \relxill\ models) for these flares is
discussed further in Section \ref{sec_spin}. Potential evidence for strong reflection
during bright flares has also been seen from \integral\ observations of this outburst
(\citealt{Roques15, Natalucci15}). We stress, though, that despite any degeneracy
between these parameters, the variations in both $E_{\rm{cut}}$ and $R_{\rm{disk}}$
are significant; if we try to force one of these two parameters to be the same for each
of the flux states and only allow the other to vary, the fits are significanty worse
($\Delta\chi^{2} > 80$ for four fewer free parameters). While the absolute values
themselves are somewhat model dependent, the trend of increasing $R_{\rm{disk}}$
with increasing flux appears to be robust to such issues.

For the disk reflection fraction to vary in such a manner, the solid angle subtended
by the disk as seen by the X-ray source must decrease as the observed flux
decreases. A few potential scenarios could produce such behaviour: (1) the disk
itself could evolve (\eg\ truncate) such that it genuinely covers a smaller solid
angle at lower fluxes; (2) the corona could evolve and vary its location/size, such
that the degree of gravitational lightbending is reduced; (3) the corona could
alternately vary its velocity, such that the beaming away from the disk is increased.
While some combination of these three effects is of course possible, and probably
even likely should the flares be related to jet ejection events, from a practical
standpoint their individual effects on the observed reflection emission are rather
similar (\citealt{Dauser13, Fabian14}). Therefore, in order to investigate the potential
geometric evolution without introducing further parameter degeneracies, we also
modify our basic model to consider two limiting scenarios representing the first 2 of
these 3 possibilities (Models 2 and 3, respectively), now making use of the fact that
given a combination of black hole spin, inner disk radius and X-ray source height,
\relxilllp\ can self-consistently compute the expected $R_{\rm{disk}}$ from the
lamppost geometry.\footnote{Models that can also self-consistently compute
$R_{\rm{disk}}$ for an X-ray source with a vertical outflow velocity are under
development, but are not yet ready for publication, limiting our ability to test the third
scenario of a variable source velocity.} First, we assume that the corona remains
static and the disk progressively truncates as the flux decreases, and second, we
assume that the disk remains static and the corona progressively moves away from
the disk (note that this could relate to either a physical movement of the corona, or a
vertical expansion of the electron cloud).

In the truncation scenario (which we call Model 2), we therefore allow the inner radius
of the disk (\rin) to vary, although given the results above we assume that during the
flares the disk does reach the ISCO, and we link $h$ across all flux states. For the
lower flux states, \rin\ is computed in units of \risco, so that we can ensure that \rin\ 
$\geq$ \risco, as simulations find that emission from within the ISCO is negligible
(\eg\ \citealt{Shafee08, Reynolds08}). In the dynamic corona scenario (Model 3), we
assume that the disk reaches the ISCO for all fluxes, and instead vary $h$. The
results from these two scenarios are presented in Table \ref{tab_param_fix}; we
focus only on the key \relxilllp\ parameters as the parameters for the other
components generally remain similar to Model 1.

\begin{figure}
\hspace*{-0.5cm}
\epsscale{1.15}
\plotone{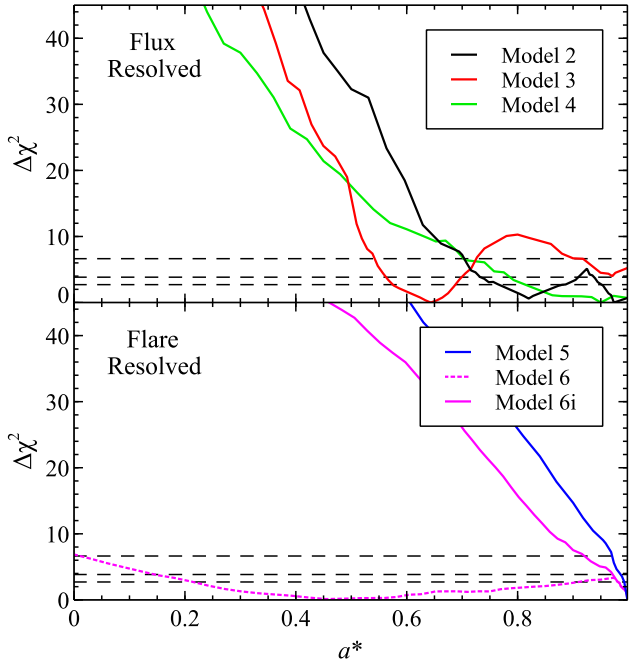}
\caption{$\Delta\chi^{2}$ confidence contours for the black hole spin for our
relativistic disk reflection models computed with a self-consistent lamppost
geometry. The top panel shows the models for our flux resolved analysis (Models
2--4, see Section \ref{sec_res}), while the bottom panel shows the models for our
analysis of the strongest six flares individually (Models 5--6; see Section
\ref{sec_flares}). For Model 6, we show both the contour calculated with no
constraint on the inclination (dotted) and with the inclination constrained to $i \geq
50$\,\deg\ (solid; Model 6i) to match the estimates for the orbital plane of the
binary system. The dashed horizontal lines indicate the 90, 95 and 99\%
confidence limits for one parameter of interest.
}
\label{fig_spin}
\end{figure}

\begin{figure}
\hspace*{-0.5cm}
\epsscale{1.15}
\plotone{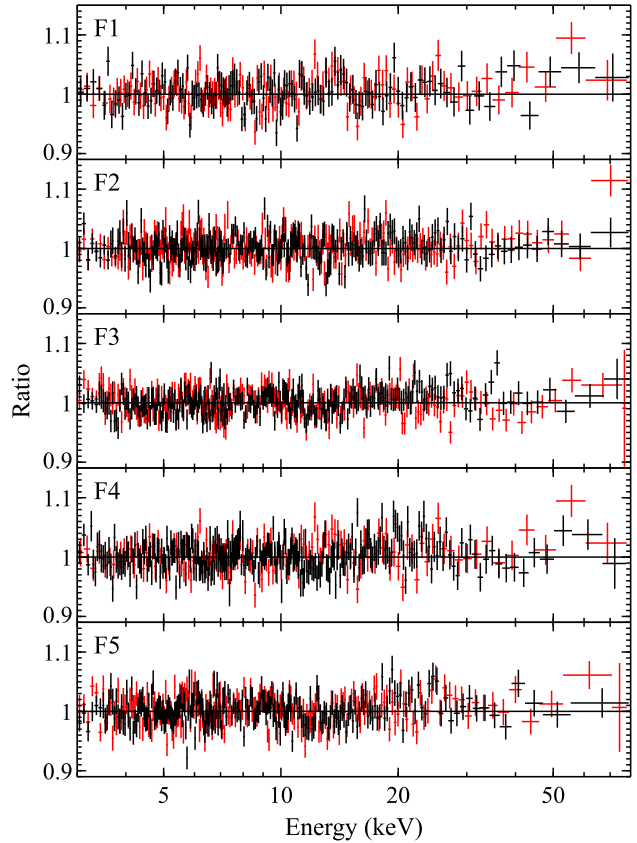}
\caption{Data/model residuals for the truncating disk model with the thermal disk
emission included from our flux-resolved analysis (Model 4; see section
\ref{sec_res}). For each of the flux states, FPMA data are shown in black, and
FPMB in red. As before, the data have been further rebinned for visual clarity.
}
\label{fig_flux_rat}
\end{figure}

Both of these scenarios provide reasonable fits to the data (\chisq/DoF =
10656/10313 and 10675/10312 for Models 2 and 3, respectively), although the
truncation scenario formally provides the better fit, and both are worse fits than
Model 1 (in which $R_{\rm{disk}}$ is a free parameter) owing to the additional
physical constraints imposed. With these additional constraints, both scenarios
require the spin to be at least moderate ($a^* \gtrsim 0.6$), but above this value
the \chisq\ landscape becomes complex. Both scenarios show distinct minima that
provide similarly good fits (different by $\Delta\chi^{2} < 5$ in both cases) at a high
spin value and at a more moderate spin value (see Figure \ref{fig_spin}). For the
truncating disk scenario (Model 2) the high spin solution ($a^* \sim 0.97$) is
marginally preferred over the lower spin solution ($a^* \sim 0.82$), while for the
dynamic corona scenario (Model 3) the lower spin solution ($a^* \sim 0.65$) is
marginally preferred over the high spin solution ($a^* \sim 0.97$), perhaps indicating
that even when allowing the height of the X-ray source to vary the data still want an
evolution in the inner radius of the disk at some level. In both of these scenarios, we
present the results for the high spin solution (which in the latter case gives a fit of
\chisq/DoF = 10678/10312) as our subsequent modeling of the individual flares
strongly suggests the black hole in \v404\ is rapidly rotating (see Section
\ref{sec_flares}). However, for completeness, we also present the parameter
constraints for the lower-spin solutions in Appendix \ref{app_lowspin}; where the
solutions are not the global best fit, errors are calculated as $\Delta\chi^{2} = 2.71$
around the local $\chi^{2}$ minimum. Separating out the solutions in this manner
also allows the evolution required in $r_{\rm{in}}$ and $h$ to be more clearly seen,
as both of these parameters are scaled by the spin in our model implementation. As
expected, we see that either the inner radius of the accretion disk moves outwards
(Model 2), or the source height moves upwards (Model 3), as the flux decreases. In
Model 2 the inner disk radius evolves from the ISCO (assumed) out to
$\sim$2.5\,\risco, and in Model 3 the source height evolves from $\sim$4 to
$\sim$6\,\rh. 

% For $a^* \sim 0.97$, \risco\ $\sim$ 1.5\,\rg, and \rh\ $\sim$ \rg.

Finally, although the X-ray emission in the \nustar\ bandpass is clearly dominated
by a hard, high-energy continuum and reprocessed emission, we also test for the
presence of any thermal emission from an accretion disk. As the self-consistent
evolutionary scenario that formally provides the best fit, we focus on the truncating
disk scenario, and modify our model for the intrinsic emission from \v404\ to
include a multi-color blackbody accretion disk (Model 4), using the \diskbb\ model
(\citealt{diskbb}). In the \xspec\ model outlined in Section \ref{sec_mod}, we thus
update the source term to be \tbabs$_{\rm{src}}$ $\times$ ( \diskbb\ $+$ \relxilllp\ ).
$R_{\rm{disk}}$ is still calculated self-consistently in the lamppost geometry from
$a^*$, $h$ and \rin. During the fitting process for this model, we found that the
\diskbb\ component only makes a significant contribution during the highest flux
state (F5), and so fixed its normalization to zero for F1--4.

\begin{figure}
\hspace*{-0.5cm}
\epsscale{1.15}
\plotone{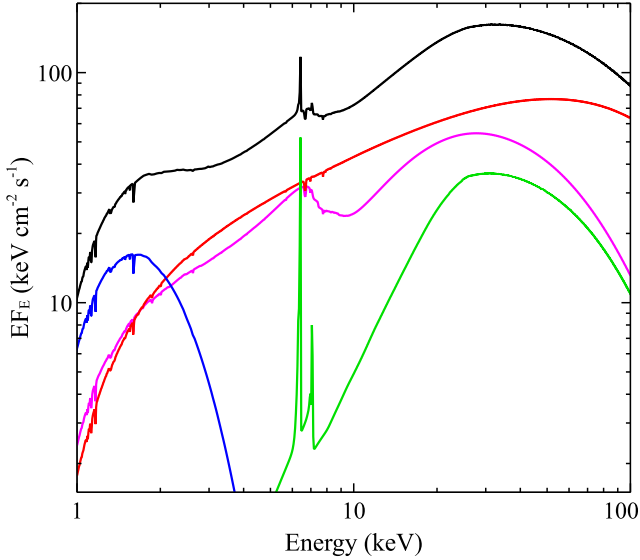}
\caption{The best-fit disk reflection model obtained for the flare spectrum (F5)
from Model 4 in our flux-resolved analysis (the truncating disk model with the
thermal disk emission included). The total model is shown in black, and the
relative contributions from the accretion disk (blue), the high-energy powerlaw
tail (red), the disk reflection (magenta) and the distant reflection (green) are also
shown.}
\vspace{0.3cm}
\label{fig_flaremod}
\end{figure}

This model provides a good fit to the data, with \chisq/DoF = 10581/10311 (\ie\ an
improvement of $\Delta\chi^{2}$ = 75 for two additional free parameters over
Model 2). We do not tabulate the parameter values, as the vast majority have not 
varied significantly from the values presented for Model 2 in Table
\ref{tab_param_fix}, but a few key parameters are worth highlighting individually.
The best fit disk temperature for the average flare spectrum is
$0.41^{+0.10}_{-0.07}$\,keV, such that this component only contributes close to the
lower boundary of the \nustar\ bandpass. However, this temperature is similar to
values reported from X-ray observatories with coverage extending to lower energies
throughout this outburst (\eg\ \citealt{Radhika16}, Rahoui et al 2016,
\textit{submitted}). The inclusion of this additional continuum component at the lower
end of the \nustar\ bandpass allows the high energy powerlaw continuum to take on
a harder photon index ($\Gamma = 1.32 \pm 0.01$), and subsequently a lower
energy cutoff ($E_{\rm{cut}} = 75 \pm 4$\,keV), such that this primary continuum
emission exhibits stronger curvature in the \nustar\ band. In turn, this allows a slightly
lower reflection fraction ($R_{\rm{disk}} = 2.5$), with the source height increasing
slightly to $h = 2.5^{+0.5}_{-0.1}$\,\rh. The black hole spin remains high, with $a^* >
0.82$ (and noteably the \chisq\ contour only displays a single solution; see Figure
\ref{fig_spin}). The data/model ratios for the five flux states are shown in Figure
\ref{fig_flux_rat} for this model, and the best-fit model along with the relative
contributions of the various emission components are shown in Figure
\ref{fig_flaremod} for the highest flux state (F5).

It is worth noting that all these flux-resolved models have returned inclinations for
the inner disk of $\sim$30\deg. This inclination would mark a large difference
between the inclination of the inner disk and the orbital plane, which the latest optical
studies during quiescence have estimated to be $i_{\rm{orb}} \sim 65$\deg\
(\citealt{Khargharia10}), with literature estimates covering a range from 50--75\deg\
(\citealt{Shahbaz94, Shahbaz96, Sanwal96}). While evidence of misalignment
between the inner and outer regions of the disk has been seen in other sources, \eg\
Cygnus X-1 (\citealt{Tomsick14, Walton16cyg}), a difference this large would likely
be unphysical (\eg\ \citealt{Fragos10, Nealon15}). We will return to this issue in the
following section.

\subsection{Individual Flares}
\label{sec_flares}

\begin{figure}
\hspace*{-0.5cm}
\epsscale{1.15}
\plotone{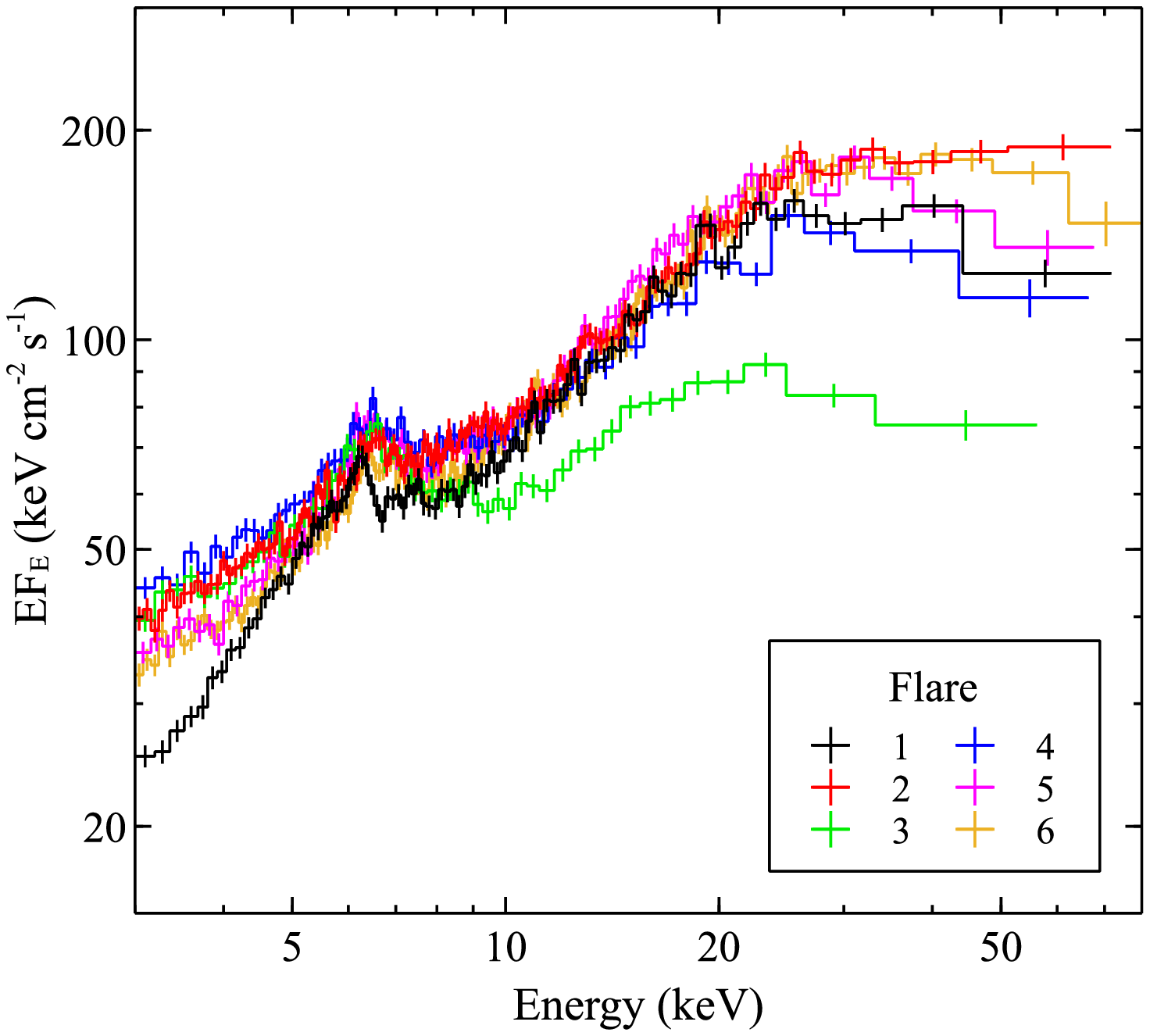}
\caption{The X-ray spectra extracted from the six major flares highlighted in Figure
\ref{fig_lcHR} (Flares 1--6 shown in black, red, green, blue, magenta and orange,
respectively). As with Figure \ref{fig_5lvl}, only the FPMA data are shown for clarity,
and the data have been unfolded through a constant and rebinned for visual
purposes. While the flares all show similar broadband hardness ratios (Figure
\ref{fig_hri}), there are clear differences between them. For example, Flare 1
(black) shows a harder spectrum at lower energies, and Flare 3 (green) shows a
softer spectrum than the rest.
}
\label{fig_allflares}
\end{figure}

We also investigate a number of the individual flares, focusing on the six that reach
or exceed $\sim$10,000\,\ctps\ (labeled in Figure \ref{fig_lcHR}). Following the
reduction procedure used in section \ref{sec_avflares}, we extracted \nustar\
spectra for each of these six flares individually. These spectra are shown in Figure
\ref{fig_allflares}. While they are all reasonably similar, as suggested by their
similar broadband hardness ratios (Figure \ref{fig_hri}), there are also obvious
differences between them, so there is still some clear averaging of different states
in our flux-resolved analysis. For example, the first flare has a harder spectrum at
lower energies than the subsequent flares, and the third flare has a softer
spectrum than the rest over the \nustar\ bandpass.

\begin{figure}
\hspace*{-0.5cm}
\epsscale{1.15}
\plotone{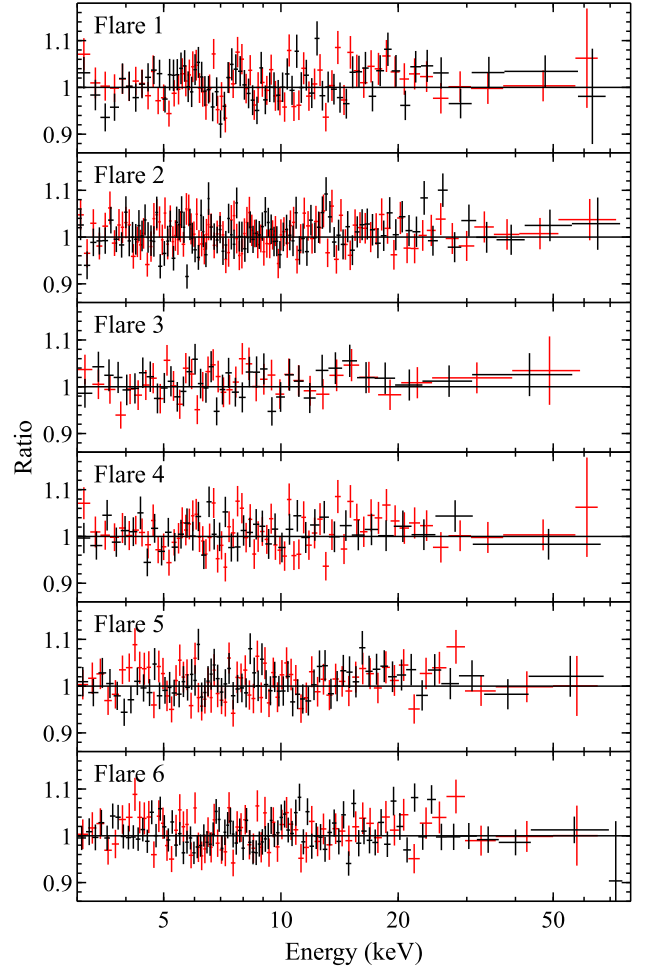}
\caption{Data/model residuals for the lamppost reflection model with the thermal
disk emission included from our flare-resolved analysis (Model 6; see section
\ref{sec_flares}). Again, for each of the flares the FPMA data are shown in black
and the FPMB data in red, and the data have been further rebinned for visual
clarity.}
\label{fig_allflares_rat}
\end{figure}

\begin{table*}
  \caption{Results obtained for the free parameters in the lamppost reflection
  models constructed for the joint fits to the individual flare spectra (Models 5
  and 6). For Model 6, the high-spin solution is given.}
  \vspace{-0.25cm}
\begin{center}
\begin{tabular}{c c c c c c c c c c}
\hline
\hline
\\[-0.1cm]
Model Component & \multicolumn{2}{c}{Parameter} & Global & \multicolumn{6}{c}{Flare} \\
\\[-0.15cm]
& & & & 1 & 2 & 3 & 4 & 5 & 6 \\
\\[-0.2cm]
\hline
\hline
\\[-0.1cm]
\multicolumn{10}{c}{Model 5: lamppost only} \\
\\[-0.1cm]
\tbabs$_{\rm{src}}$ & \nh\ & [$10^{22}$ cm$^{-2}$] & & $3.3^{+0.6}_{-0.8}$ & $<0.2$ & $<0.3$ & $<0.2$ & $<0.1$ & $<0.1$ \\
\\[-0.2cm]
\relxilllp\ & $\Gamma$ & & & $1.22^{+0.06}_{-0.12}$ & $1.44^{+0.01}_{-0.03}$ & $1.63^{+0.08}_{-0.12}$ & $1.52^{+0.03}_{-0.04}$ & $1.28^{+0.02}_{-0.03}$ & $1.58^{+0.01}_{-0.03}$ \\
\\[-0.2cm]
& $E_{\rm{cut}}$ & [keV] & & $50^{+6}_{-10}$ & $210 \pm 30$ & $47 \pm 4$ & $94^{+12}_{-15}$ & $60^{+4}_{-6}$ & $140^{+20}_{-10}$ \\
\\[-0.2cm]
& $a^*$ & & $>0.99$ \\
\\[-0.2cm]
& $i$ & [\deg] & $42\pm2$ \\
\\[-0.2cm]
& $h$ & \rh\ & & $<2.2$ & $<2.4$ & $4.0^{+3.5}_{-1.0}$ & $2.7^{+1.6}_{-0.3}$ & $<2.1$ & $8.0^{+3.3}_{-0.9}$ \\
\\[-0.2cm]
& $A_{\rm{Fe}}$ & [solar] & $2.9^{+0.3}_{-0.6}$ \\
\\[-0.2cm]
& $\log\xi$ & $\log$[\ergcmps] & & $3.2 \pm 0.1$ & $3.5^{+0.2}_{-0.1}$ & $3.1^{+0.2}_{-0.1}$ & $3.4^{+0.2}_{-0.1}$ & $3.36^{+0.07}_{-0.05}$ & $2.2 \pm 0.1$  \\
\\[-0.2cm]
& $R_{\rm{disk}}$\tmark[a] & & & $5.3$ & $4.4$ & $2.3$ & $3.4$ & $5.3$ & $1.5$ \\
\\[-0.2cm]
& Norm & & & $4.2^{+0.3}_{-0.5}$ & $4.4^{+0.9}_{-0.7}$ & $1.1^{+0.3}_{-0.4}$ & $2.1^{+0.5}_{-1.1}$ & $3.9^{+0.3}_{-1.2}$ & $1.3 \pm 0.2$ \\
\\[-0.2cm]
\xstar$_{\rm{abs}}$ & $\log\xi$ & $\log$[\ergcmps] & $5.3^{+0.4}_{-0.3}$ \\
\\[-0.2cm]
& \nh\ & [$10^{21}$ cm$^{-2}$] & & $43^{+12}_{-10}$ & $<1$ & $<3$ & $<7$ & $<3$ & $<12$ \\
\\[-0.2cm]
\xillver\ & Norm & & & $0.22 \pm 0.04$ & $0.42^{+0.09}_{-0.08}$ & $0.25 \pm 0.06$ & $0.34^{+0.04}_{-0.07}$ & $0.45 \pm 0.05$ & $0.40 \pm 0.04$ \\
\\[-0.2cm]
\hline
\\[-0.1cm]
\chisq/DoF & & & 6039/5859 & \\
\\[-0.2cm]
\hline
\hline
\\[-0.1cm]
\multicolumn{10}{c}{Model 6: lamppost with disk emission} \\
\\[-0.1cm]
\tbabs$_{\rm{src}}$ & \nh\ & [$10^{22}$ cm$^{-2}$] & & $5.1^{+1.0}_{-1.1}$ & $2.3^{+0.9}_{-0.6}$ & $3.6^{+1.4}_{-1.3}$ & $2.2^{+1.5}_{-1.3}$ & $3.3^{+1.1}_{-1.0}$ & $3.8^{+0.8}_{-0.9}$ \\
\\[-0.2cm]
\diskbb\ & $T_{\rm{in}}$ & [keV] & $0.49 \pm 0.04$ \\
\\[-0.2cm]
& Norm & [$10^5$] & & $1.5^{+1.3}_{-0.7}$ & $1.9^{+1.5}_{-1.1}$ & $3.4^{+2.9}_{-1.5}$ & $2.2^{+1.4}_{-1.1}$ & $2.7^{+2.2}_{-1.2}$ & $2.8^{+1.9}_{-1.2}$ \\
\\[-0.2cm]
\relxilllp\ & $\Gamma$ & & & $<1.04$\tmark[b] & $1.27^{+0.04}_{-0.03}$ & $1.29^{+0.12}_{-0.08}$ & $1.33^{+0.07}_{-0.06}$ & $<1.12$\tmark[b] & $1.22^{+0.06}_{-0.04}$ \\
\\[-0.2cm]
& $E_{\rm{cut}}$ & [keV] & & $40^{+3}_{-2}$ & $113^{+16}_{-28}$ & $32^{+5}_{-4}$ & $60^{+12}_{-9}$ & $42^{+5}_{-3}$ & $68^{+8}_{-6}$ \\
\\[-0.2cm]
& $a^*$ & & $>0.98$ \\
\\[-0.2cm]
& $i$ & [\deg] & $52^{+2}_{-3}$ \\
\\[-0.2cm]
& $h$ & \rh\ & & $<2.3$ & $<3.8$ & $<3.6$ & $<3.5$ & $<2.4$ & $7.5^{+8.7}_{-1.9}$ \\
\\[-0.2cm]
& $\log\xi_{\rm{disk}}$ & $\log$[\ergcmps] & & $3.5 \pm 0.1$ & $3.8 \pm 0.1$ & $3.4 \pm 0.1$ & $3.6 \pm 0.1$ & $3.6 \pm 0.1$ & $3.5 \pm 0.1$ \\
\\[-0.2cm]
& $A_{\rm{Fe}}$ & [solar] & $5.0^{+0.7}_{-0.4}$ \\
\\[-0.2cm]
& $R_{\rm{disk}}$\tmark[a] & & & 5.3 & 3.9 & 4.1 & 4.1 & 5.3 & 1.6 \\
\\[-0.2cm]
& Norm & & & $3.5^{+0.2}_{-0.8}$ & $3.1^{+2.0}_{-1.7}$ & $1.9^{+1.0}_{-0.9}$ & $2.7^{+1.4}_{-1.3}$ & $3.5^{+0.2}_{-1.2}$ & $1.0^{+0.4}_{-0.2}$ \\
\\[-0.2cm]
\xstar$_{\rm{abs}}$ & $\log\xi$ & $\log$[\ergcmps] & $5.7 \pm 0.2$ \\
\\[-0.2cm]
& \nh\ & [$10^{21}$ cm$^{-2}$] & & $88^{+4}_{-3}$ & $<6$ & $<16$ & $<30$ & $<23$ & $19^{+17}_{-11}$ \\
\\[-0.2cm]
\xillver\ & Norm & & & $0.13 \pm 0.05$ & $0.22^{+0.09}_{-0.08}$ & $0.13 \pm 0.07$ & $0.19 \pm 0.09$ & $0.29 \pm0.06$ & $0.26^{+0.06}_{-0.03}$ \\
\\[-0.2cm]
\hline
\\[-0.15cm]
\chisq/DoF & & & 5906/5852 & \\
\\[-0.2cm]
\hline
\\[-0.15cm]
$F_{3-79}$\tmark[c] & \multicolumn{2}{c}{[$10^{-8}$\,\ergpcmsqps]} & & $50.7 \pm 0.6$ & $62.4 \pm 0.6$ & $34.4 \pm 0.5$ & $50.6 \pm 0.6$ & $57.0 \pm 0.6$ & $60.0 \pm 0.4$ \\
\\[-0.2cm]
\hline
\hline
\end{tabular}
\vspace{-0.2cm}
\label{tab_flares}
\end{center}
$^a$ For these models, $R_{\rm{disk}}$ is calculated self-consistently in the
lamppost geometry from $a^*$ and $h$. As it is not a free
parameter, errors are not estimated. \\
$^b$ The RELXILLLP model is only calculated for $\Gamma \geq 1$. \\
$^c$ Average flux in the 3--79\,keV bandpass.
\vspace{0.5cm}
\end{table*}

We performed a joint fit of each of these flare spectra with the lamppost model
discussed in sections \ref{sec_mod} and \ref{sec_res} (excluding the photoionised
emission component, which makes no contribution to the flux-resolved fits at high
fluxes). As with our flux-resolved analysis, we link the black hole spin, iron
abundance, accretion disk inclination, and ionization state of the photoionized
absorption across all the flares. Additionally, for the distant reprocessor, we fix the
shape parameters (iron abundance, slab inclination) to the values found in the
flux-resolved work. We also assume that the disk extends to the ISCO and again
compute the reflection fraction self-consistently assuming a lamppost geometry.
Finally, given the results presented in Section \ref{sec_res}, we fit the lamppost
model both with and without an accretion disk contribution, again using the
\diskbb\ model. While fitting the model with the \diskbb\ component, we found
the disk temperatures to be consistent among all the flares, and so linked this
parameter across the datasets for simplicity. The results obtained with both these
models (Models 5 and 6, respectively) are presented in Table \ref{tab_flares}.

The pure lamppost model (Model 5) fits the data well, with with \chisq/DoF =
6039/5859. The spin is constrained to be very high, $a^* > 0.99$ (see Figure
\ref{fig_spin}), and there is a slight increase in the inclination inferred for the
inner disk; while the flux-resolved analysis typically found $i \sim 30$\deg, here
we find $i \sim 40$\deg. We find that the first flare shows stronger absorption
than the subsequent flares, both in terms of the neutral and the ionized
absorption components. The former results in the harder spectrum seen from
this flare at lower energies, and there is a clear absorption line from ionized
iron at $\sim$6.7\,keV produced by the latter, similar to that reported by
\cite{King15v404}, which is not seen in any of the subsequent flares. As with
the flux-resolved analysis, we find that during the flares the height inferred for
the X-ray source is very close to the black hole, always within $\sim$10\,\rg.

The model including the disk emission (Model 6) again provides a substantial
improvement over the basic lamppost model, resulting in an outstanding fit to the
data (\chisq/DoF = 5906/5852, \ie\ an improvement of $\Delta\chi^{2}$ = 133 for 7
additional free parameters). We show the data/model ratios for the individual
flares with this model in Figure \ref{fig_allflares_rat}. The disk temperature is again
similar to that reported by lower energy missions, $T_{\rm{in}} \sim 0.5$\,keV, and
as before we see that the inclusion of this emission allows the high-energy
continuum to take on a harder form, subsequently resulting in lower-energy cutoffs.
The neutral absorption inferred also increases to compensate for this additional
low-energy continuum emission. 

With regards to the black hole spin, we again find a situation in which two 
solutions exist that provide statistically equivalent fits (separated by $\Delta\chi^{2}
< 1$; see Figure \ref{fig_spin}): one at high spin ($a^* > 0.98$) which is marginally
preferred, and another broad, local minimum at a more moderate spin ($a^* \sim
0.5$). In this case, the dual solutions are related to a significant degeneracy between
the spin and the disk inclination, resulting from the combination of the additional
continuum component, and the lower total S/N utilized in these fits (these data
represent $\sim$80\% of the exposure from which the F5 spectrum considered in the
previous section is extracted).

For the best-fit, high spin solution we find that the inclination has further increased to
$i \sim 52$\deg, which is similar to the estimates for the orbital inclination of the
system ($i_{\rm{orb}} \sim 50$--75\deg; \eg\ \citealt{Shahbaz94, Khargharia10}). In
contrast, for the more moderate spin solution we find that the associated inclination is
$<20$\deg. This would imply an even more extreme disk warp than the flux-resolved
analysis, which we deem unphysical. This degeneracy between the spin and the
inclination is distinct from the traditional sense of a parameter degeneracy, in which
two parameters are correlated such that any value of one can be made acceptable by
adjusting the other; rather there are two solutions that are acceptable in distinct areas
of parameter space. We therefore present the results from the high-spin solution in
Table \ref{tab_flares}, although again the parameter constraints for the lower-spin
solution are presented in Appendix \ref{app_lowspin}, and re-calculate the confidence
contour for the black hole spin with the inclination constrained to be $i \geq
\ilim$\deg\ for this model (which we refer to as Model 6i; see Figure \ref{fig_spin}) in
order to ensure a reasonable agreement between the inner and outer disk. This
constraint strongly requires a rapidly rotating black hole. We also assess the degree
to which the assumed geometry is driving the spin constraint in this scenario by
relaxing the requirement that $R_{\rm{disk}}$ is set self-consistently and allowing this
to vary as a free parameter for each of the 6 flares (but keeping the $i \geq 45$\deg\
constraint). Although the constraint on the spin is naturally looser, we still find that $a^*
> 0.7$ and the constraints on $R_{\rm{disk}}$ are all consistent with the values
presented in Table \ref{tab_flares}. If we exclude unphysically large disk warps, a
rapidly rotating black hole is still required regardless of any additional geometric
constraints.

\begin{figure*}
\hspace*{-0.5cm}
\epsscale{1.1}
\plotone{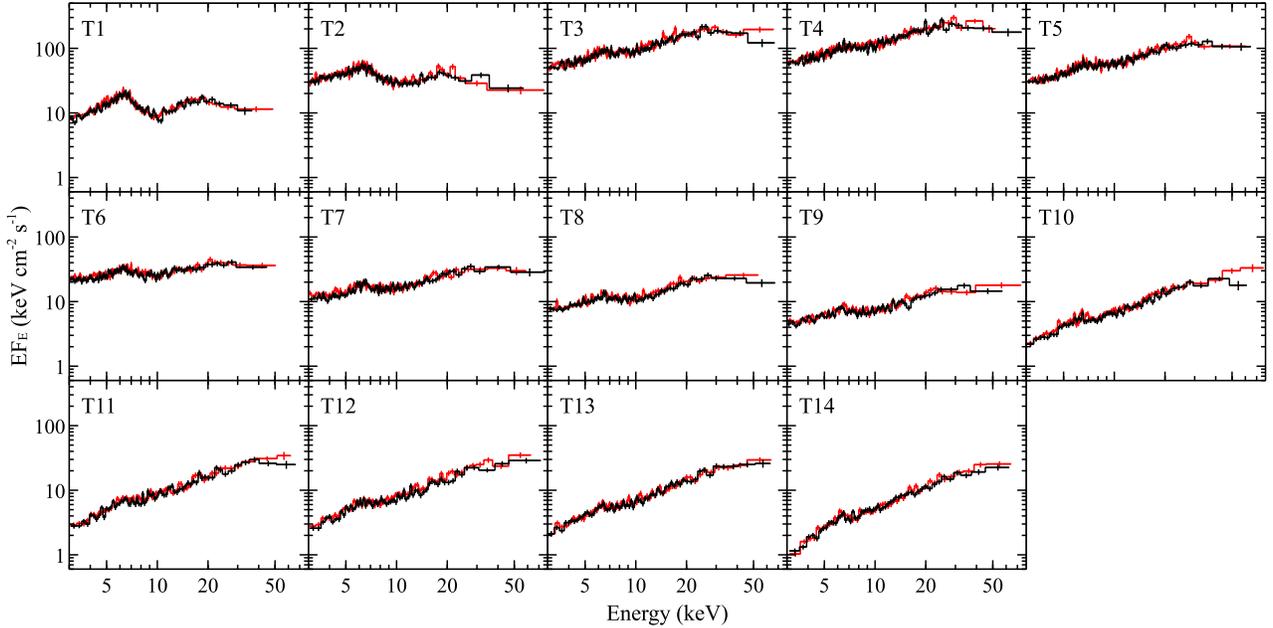}
\caption{The 14 time-resolved X-ray spectra extracted across the evolution of flare 4
(labeled T1--14). As before, the data have been unfolded through a constant and
rebinned for visual purposes, and the FPMA and FPMB data are shown in black and
red, respectively. Significant spectral changes are seen as the source flares and
then decays.
}
\vspace{0.3cm}
\label{fig_flare4spec}
\end{figure*}

The range of heights inferred for the X-ray source remains similar to the pure
lamppost case. However, one issue of note with this model is that the iron
abundance has increased to $A_{\rm{Fe}}$/solar $\sim$ 5 (for both solutions), in
order to compensate for the harder irradiating continuum and reproduce the
observed line flux. All our previous models had typically found $A_{\rm{Fe}}$/solar
$\sim$ 2--3, which is similar to the iron abundance of $A_{\rm{Fe}}$/solar $\sim$
2 found for the companion star by \cite{GonzHern11}. While this is certainly not
always the case (\eg\ \citealt{ElBatal16}), similarly high iron abundances have also
been reported for a few other Galactic BHBs observed by  \nustar\ when using the
\xillver\ based family of reflection models (\eg\ \citealt{Parker16, Walton16cyg,
Fuerst16gx}). The abundance inferred may be dependent on the reflection code
utilized; \cite{Walton16cyg} note that for the Galactic binary Cygnus X-1, the iron
abundances obtained with the \xillver\ family of reflection models are generally a
factor of $\sim$2 larger than those obtained with the \reflionx\ (\citealt{reflion})
family of models (see also \citealt{Miller15}). Should the iron abundance here be
systematically overpredicted by a similar factor, this would bring the abundance
derived back down to $A_{\rm{Fe}}$/solar $\sim$ 2.5, which would again be similar
to that reported by \cite{GonzHern11}. However, we stress that the key results
obtained here do not strongly depend on this issue. If we fix the iron abundance to
$A_{\rm{Fe}}$/solar = 2 in Model 6, the fit worsens slightly (but is still excellent,
\chisq/DoF = 5933/5853). The spin is strongly constrained to be very high ($a^* >
0.997$), and the requirement for small source heights further tightens. The most
noteable change is that the best-fit inclination further increases to $i \sim 60$\deg,
which is still in good agreement with the range estimated for the orbital plane.

The 3--79\,keV fluxes observed from these spectra are also given in Table
\ref{tab_flares}. However, the average count rates during the periods from which
these spectra are extracted are obviously significantly lower than the peaks of the
flares. Assuming a similar spectral form, scaling these fluxes up to the peak
incident count rates observed during these flares -- as determined from lightcurves
with 1s time bins -- corresponds to peak 3--79\,keV fluxes ranging from 0.8--2.0
$\times 10^{-6}$\,\ergpcmsqps. For a 10\,\msun\ black hole at a distance of
2.4\,kpc, these fluxes equate to 3--79\,keV luminosities of $\sim$0.4--1.0\,\ledd\
(where \ledd\ = $1.4 \times 10^{39}$ \ergps\ is the Eddington luminosity). The
bolometric fluxes observed from the \diskbb\ component in these spectra, which
assumes a thin disk as described by \cite{Shakura73}, equate to disk luminosities
of $\sim$0.1\,\ledd\ (assuming the disk is viewed close to $i \sim 60$\deg).
Temperatures of $T_{\rm{in}} \sim 0.5$ keV are not unreasonable for such
luminosities (\eg\ \citealt{Gierlinski04, MReynolds13}). Assuming these fluxes also
scale up during the peaks of the flares, the peak disk fluxes would equate to
luminosities of $\sim$0.3--0.5\,\ledd.

\subsection{Evolution Across Flare 4}
\label{sec_flare4}

As the final component of our analysis in this work, we track the evolution of the
spectrum across one of the major flares considered in Section \ref{sec_flares}.
We focus on Flare 4 (see Figure \ref{fig_lcHR}), as this is followed by a relatively
long, uninterrupted period of low absorption (as determined by our analysis in
Section \ref{sec_sel}). As such, we should have a relatively clean view of the flare
and its subsequent decline. In order to track the evolution of the spectrum,
we split the data into bins with 40\,s duration, and extracted spectra from each,
again following the method outlined in Section \ref{sec_red}. While significant 
variability obviously occurs on shorter timescales (\eg\ Gandhi et al. in preparation),
40\,s duration was found to offer a good balance between retaining good time
resolution and the need for reasonable S/N in the individual spectra. We start
immediately prior to the flare, and continue until the point that the observed count
rate (as averaged over 40\,s) starts to rise again after the decline of the flare,
resulting in 14 time-resolved spectra (per FPM) in total (hereafter T1--14). These
spectra are shown in Figure \ref{fig_flare4spec}.

There are too many datasets to undertake a joint analysis of all the data, so we fit
the data from each of the time bins individually, using the same lamppost-based
model utilized in our joint analysis of the major flares observed (Section
\ref{sec_flares}). Specifically, we use the model that includes the thermal disk
emission (Model 6). However, the average good exposure time per FPM is only
$\sim$11\,s per bin (being higher for lower flux bins and vice versa, owing to the
instrumental deadtime; \citealt{NUSTAR}), so the S/N per time bin is relatively low.
We therefore limit ourselves to considering only a few key free parameters when
fitting each of these datasets. As there is no evidence for ionized iron absorption
during this flare (only an upper limit is obtained on the column for this component
during flare 4, see Table \ref{tab_flares}), we exclude the \xstar\ absorption
component from our analysis in this section. Furthermore, we fix all the remaining
global parameters (black hole spin, disk inclination, iron abundance and disk
temperature) to the best fit values presented for Model 6 in Table \ref{tab_flares}.
We also fix the ionization parameter to the value obtained in our flux-resolved
analysis (see Table \ref{tab_param}), based on the average count rate in that time
bin, thus ensuring that the ionization increases as the flux increases. Finally, we are
not able to simultaneously constrain both the inner radius of the disk and the height
of the X-ray source, so we initially fix the latter at the best-fit obtained for this flare
in our flare-resolved analysis ($h = 2.5$\,\rh). The free parameters allowed to vary
for each of the time-resolved datasets are therefore the (source intrinsic) neutral
absorption column, the photon index and high-energy cutoff of the powerlaw
continuum, the inner radius of the disk, and the normalizations of the various
emission components. As before, the reflection fraction $R_{\rm{disk}}$ is
calculated self-consistently from the spin, source height and inner radius of the
disk in the lamppost geometry, which helps to constrain \rin\ in these fits.

\begin{figure}
\hspace*{-0.5cm}
\epsscale{1.15}
\plotone{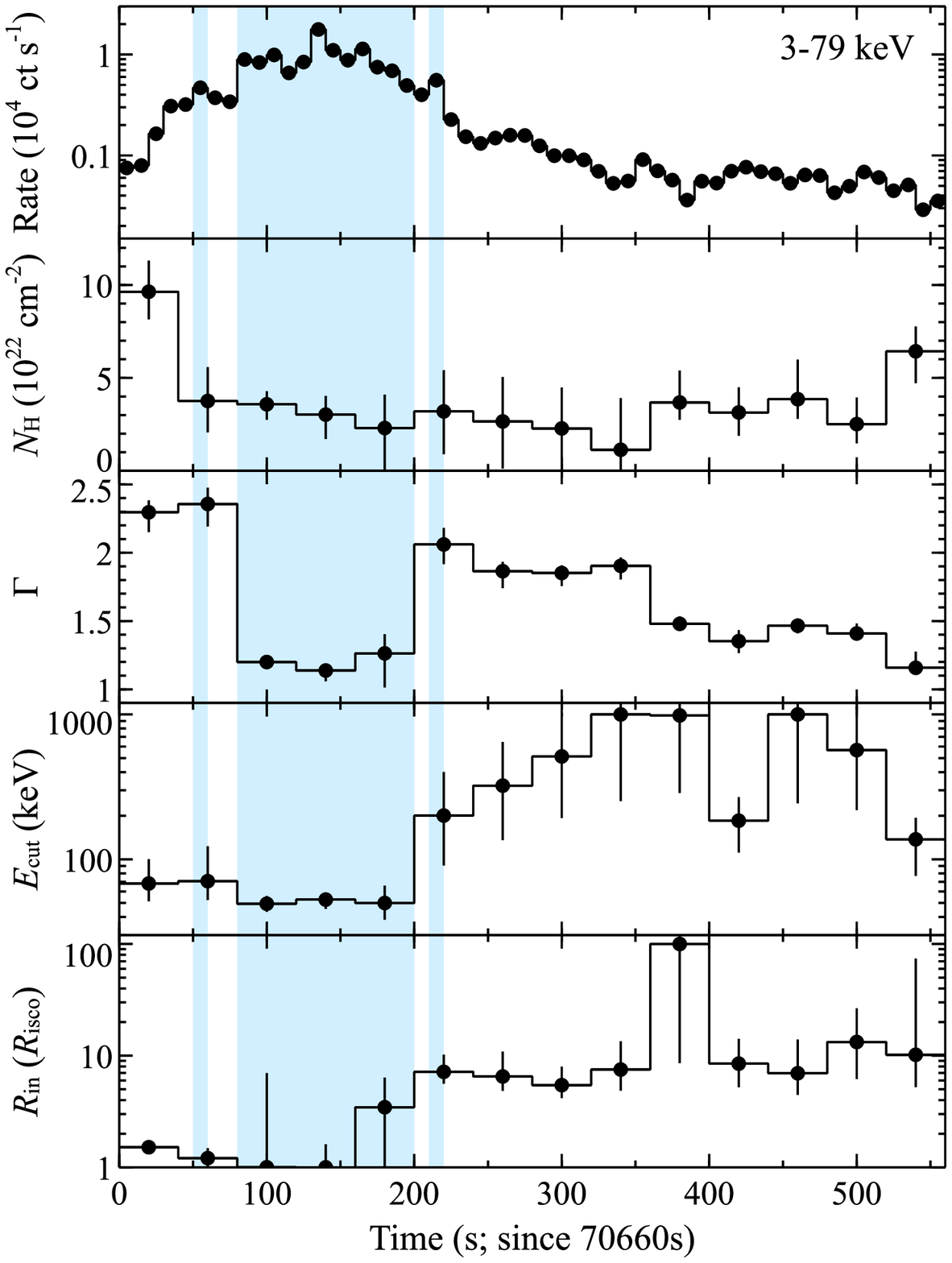}
\caption{The results for the key lamppost model parameters obtained with our
time-resolved spectral analysis of Flare 4, in this case allowing the inner radius of the
disk to vary while holding the height of the X-ray source constant (see text). The top
panel shows the lightcurve around flare 4 (10s bins), while the lower
panels show the evolution of the intrinsic neutral absorption column, the photon index
and high-energy cutoff of the powerlaw continuum, and the inner disk radius,
respectively (each 40s bins). The high-energy continuum hardens significantly during
the peak of the flare. In addition, both the high-energy cutoff and the inner radius of
the disk are significantly larger after the peak of the flare than before. The shaded
regions indicate periods when the count rate exceeds 4000\,\ctps, which contribute
to the Flare 4 spectrum shown in Figure \ref{fig_allflares}.
}
\vspace{0.3cm}
\label{fig_flare4res}
\end{figure}

The results for a number of the key parameters, as well as a zoom-in on the 
lightcurve of this flare, are shown in Figure \ref{fig_flare4res}, which shows a
characteristic fast rise, exponential decay profile. Aside from the first time bin, the
absorption stays relatively low and stable throughout, as expected. Prior to the flare,
the observed spectrum is relatively soft (in comparison to the spectra shown in
Figures \ref{fig_5lvl} and \ref{fig_allflares}). Then, as the source flares the spectrum
hardens significantly (reaching $\Gamma = 1.14^{+0.04}_{-0.08}$), and during the
decline it softens again before gradually becoming harder as the source fades. We
see a significant difference in the average cutoff energy before and after the flare.
Finally, we also see a significant difference in the inner radius of the disk, the key
geometry parameter in this analysis, across the evolution of the flare, being close to
the ISCO prior to and during the rise of the flare, before moving out to
$\sim$10\,\risco\ during the subsequent decline. The data are well modeled, with an
average \chisq/DoF of 1.02 (for an average of 302 DoF). As a sanity check, assuming
a disk structure of $h_{\rm{D}}/r_{\rm{D}} \sim 0.2$ (where $h_{\rm{D}}$ is the scale
height of the disk at a given radius $r_{\rm{D}}$; see Section \ref{sec_spin}), a
standard viscosity parameter of $\alpha \sim 0.1$, and that the dynamical timescale
is set by the Keplerian orbital timescale, we estimate the viscous timescale for the
disk should be $\sim$0.01\,s at a radius of 10\,\rg\ (for our best-fit spin, \risco\ $\sim$
\rg) for \v404. Significant evolution of the inner disk is therefore certainly possible
over the timescales probed here.

We also consider two additional iterations of this analysis. First, as with our
flux-resolved analysis, we also consider the case in which $h$ varies and \rin\ stays
constant, fixing the latter to the ISCO throughout. Equivalent results are obtained,
with the only difference being that $h$ increases as the flare evolves instead of \rin,
starting at $\sim$2\,\rh\ before jumping to $\sim$20\,\rh\ in the decline of the flare.
The fit statistics are very similar to the scenario in which \rin\ varies and $h$ is
constant. At least one of $h$ or \rin\ must therefore increase across the flare; in
reality the two may well evolve together. Second, we relax our assumption with
regards to the ionization of the disk. While this would be expected to increase with
increasing luminosity for a constant density, with the inner regions of the disk
evolving its density may also vary. We therefore re-fit the data with the ionization
as a further free parameter. While this increases the uncertainties on the other
parameters, the same qualitative evolution is still seen, with the main difference
being that the point at which \rin\ moves outwards occurs later in time. Broadly
speaking, the ionization of the disk does still appear to increase with increasing flux.

Finally, we note that \v404\ is known to exhibit a strong dust halo, which can produce
emission that can potentially mimic an accretion disk component, particularly when
the source is faint (\citealt{Vasilopoulos16, Beardmore16, Heinz16, Motta16v404}).
However, during this work we are largely focusing on periods when the source was
very bright. Furthermore, in the analysis presented here we find that the normalization
of the \diskbb\ component included in the model varies across Flare 4 along with the
overall flux. This is too fast for the response from dusty interstellar clouds, and so we
cannot be mistaking a dust contribution for the accretion disk in this work.

\section{Discussion}
\label{sec_dis}

We have undertaken an analysis of the first of a series of \nustar\ observations of
\v404\ taken across its recent outburst in summer 2015. This observation was
taken during the period of extreme activity from the source (see Figure
\ref{fig_longlc}). Extreme flux and spectral variability is present throughout (see
Figure \ref{fig_lcHR}), driven in part by strong and variable line-of-sight absorption,
similar to that seen in the last major outburst from this source in 1989 (\eg\
\citealt{Zycki99b}). We also see a period of intense flaring, similar to that reported
by other high-energy observatories (\eg\ \citealt{Rodriguez15v404, Natalucci15,
Roques15, King15v404, Jenke16v404}), with the source reaching observed fluxes
that correspond to its Eddington luminosity in the 3--79\,keV band in the most
extreme cases covered by \nustar. Given the strength of these flares, the ability of
\nustar\ to cleanly observe extreme count rates free of instrumental effects such
as pile-up, owing to its triggered read-out (\citealt{NUSTAR}), has been critical to
this work.

Our analysis focuses primarily on this flaring period. While the line-of-sight absorption
is often strong during this observation, as indicated by the strong edge seen at
$\sim$7\,keV in the average spectrum from the entire observation (see Figure
\ref{fig_spec_av}), the average spectrum extracted from the highest fluxes (the flare
peaks) seen during this period shows comparatively little absorption, with no strong
edge seen, and thus offers us a relatively clean view of the intrinsic spectrum from
\v404. These data show clear evidence of relativistic reflection from an accretion disk
(Figure \ref{fig_flares}), as well as reprocessing from more distant material (see also
\citealt{King15v404, Motta16v404}). We undertake a series of detailed analyses in
order to determine the relative contributions of these components, and probe the
geometry of the inner accretion flow during these flares.

First, we use these flares as a template to identify further periods of low absorption
throughout the rest of the \nustar\ observation, and undertake a flux-resolved
analysis of these data (Section \ref{sec_flux}), averaging them into five flux bins and
fitting these simultaneously with the latest self-consistent disk reflection model,
assuming a lamppost geometry (\relxilllp; \citealt{relxill}). The relative contribution of
the disk reflection decreases with decreasing flux. The evolution of the strength of 
the disk reflection implies that, on average, the solid angle subtended by the disk,
as seen by the illuminating X-ray source, decreases with decreasing flux. In turn, this
requires an evolution in the geometry of the innermost accretion flow. To minimize
parameter degeneracies we tested two limiting scenarios based on an idealized
lamppost approximation for the accretion geometry, first in which the changing solid
angle is explained with a truncating disk and a static illuminating source, and second
with a stable disk and a changing source height (resulting in a varying degree of
gravitational lightbending). The latter scenario could potentially represent either a
physical motion or a vertical expansion of the X-ray source. We note, however, that it
is possible (if not likely, as discussed below) that both the inner radius of the disk and
the height of the X-ray source could be varying simultaneously. Both of the scenarios
considered suggest that during the peaks of the flares, the average position of the
X-ray source is close to the black hole ($h \lesssim 5$\,\rg). In addition to the
high-energy powerlaw continuum and the reprocessed emission components that
dominate the majority of the \nustar\ band, we also find evidence for a weak
contribution from thermal emission from the disk in the highest flux bin, seen at the
lowest energies probed (see Figure \ref{fig_flaremod}). The lower flux data do not
show any evidence for such emission in the \nustar\ band.

Second, we undertake a joint analysis of the spectra extracted from the peaks of the
six strongest flares observed (highlighted in Figure \ref{fig_longlc}; Section
\ref{sec_flares}). We again fit the data with our lamppost disk reflection model in order
to build on our previous analysis and probe the geometry during these flares
individually. While these flares all have broadly similar spectra, there are also
differences between them (Figure \ref{fig_allflares}), so it is important to assess what
effect the averaging of different spectra inherent to our flux-resolved analysis might
have on the results obtained. Our analysis of these data with our lamppost disk
reflection model finds further support for the contribution of thermal disk emission at
the highest fluxes, and also confirms that the X-ray source is indeed close to the
black hole (within $\sim$10\,\rg) during these flares.

With the strong gravitational light bending associated with this regime resulting in an
increased fraction of the emitted flux being lost over the black hole horizon and/or
bent onto the accretion disk, the intrinsic power emitted during these flares would be
even larger than simply inferred from the observed fluxes. For the high spin solutions,
the work of \cite{Dauser14} suggests that only $\sim$20\% of the intrinsically emitted
flux should be lost over the event horizon, so the reflection fraction -- defined here to
be the ratio of the fluxes seen by the disk and by the observer -- provides a
reasonably good scaling factor between the observed and intrinsic fluxes. At the flare
peaks, we would therefore infer the hard X-ray continuum to be intrinsically $\sim$4
times brighter (on average) than observed based on our flare-resolved analysis.
However, we stress that this correction is geometry dependent, and even within the
assumed geometry depends strongly on the source height; increasing $h$ within the
formal statistical uncertainties quoted in Table \ref{tab_flares} can reduce this factor
quite substantially (by up to $\sim$40\%).

Finally, we undertake a time-resolved analysis of the evolution across one of these
major flares, focusing on flare 4 (Section \ref{sec_flare4}). Spectra are extracted every
40\,s, and fit individually with our lamppost disk reflection model. Owing to the short
exposures, the S/N in each spectrum is relatively poor. We therefore again focus on
the limiting scenarios in which the inner radius of the disk varies while the height of
the X-ray source remains constant, and vice versa (although we again stress that this
is for pragmatic reasons regarding parameter degeneracies, and that both quantities
may in reality vary together, as discussed below). In both cases, we find clear
differences before and after the peak of the flare, so at least one of these quantities
must evolve across the flare; either the disk truncates, or the height of the source
increases (Figure \ref{fig_flare4res}). During the peak of the flare, the primary
continuum is extremely hard ($\Gamma \sim 1.1$), and we also see a clear evolution
in the high-energy cutoff, which is significantly higher after the peak of the flare than it
was before.

\subsection{Jet Activity}
\label{sec_jet}

We suggest that the strong flares observed by \nustar\ mark transient jet ejection
events, with the jet becoming the source of the X-rays illuminating the disk (hence our
use of the lamppost geometry throughout our reflection modeling). There are a variety
of lines of evidence from the X-ray band alone that support this claim. In other
accreting black holes, strong X-ray flares are known to be associated with such events.
The BHB XTE\,J1550-564 is particularly notable in this respect. During its 1998
outburst, an $\sim$Eddington level flare was observed by \rxte, which triggered the
onset of super-luminal radio ejecta, and some time later ejecta were resolved from the
central point source by \chandra\ (\citealt{Corbel02, Tomsick03, Kaaret03, Steiner12}).
Such behavior has also been seen from the BHB H\,1743$-$322 (\citealt{Corbel05,
McClintock09, Steiner12h1743}), and the X-ray flux seems to be elevated in the hours
prior to many of the radio ejections from the BHB GRS\,1915+105 (\eg\
\citealt{Punsly13, Punsly16}). In addition, some X-ray flares in active galaxies also
appear to be associated with jet ejection events (\eg\ the recent flares observed from
Mrk\,335 and M81; \citealt{Wilkins15, King16m81}). The fact that the intrinsic spectrum
observed immediately prior to flare 4 in our time-resolved analysis is inferred to be
quite soft ($\Gamma \sim 2.3$) is of potential importance here, as this is very similar
to the `steep powerlaw state' identified by \citet[also referred to as the `Very High
State' in other works]{Remillard06rev}. For most LMXBs in outburst, transient jets are
launched as they flare up to $\sim$Eddington during the transition from the hard state
to the soft state, which occurs via the steep powerlaw state (\eg\ \citealt{Fender04,
Corbel04, Steiner11, Narayan12}).

In addition to this broader precedent, the nature of the X-ray spectrum observed
during these flares also supports a jet scenario. Even after accounting for the
reprocessed emission, the primary X-ray continuum is found to be extremely hard,
despite the high flux; on average we see $\Gamma \sim 1.4$, and from our
time-resolved analysis of flare 4 we see that the continuum even reaches $\Gamma
\sim 1.1$. This is not the spectrum that would be expected from an accretion flow
radiating at $\sim$Eddington, which should be dominated by emission from a
multi-color blackbody accretion disk, modified slightly by the effects of photon
advection (\eg\ \citealt{Middleton12, Middleton13nat, Straub13}). In addition, spectra
this hard (particularly in the $\Gamma \sim 1.1$ case) are difficult to produce via
Compton scattering of thermal disk photons in a standard accretion disk corona.
Strong illumination of the corona by the disk should cool the electrons and produce
a softer spectrum. The hard X-ray source would therefore be required be extremely
photon starved (\eg\ \citealt{Fabian88, Haardt93}), in which case only a very small
fraction of the disk emission would be scattered into the hard X-ray continuum, or
some other process must serve to counteract the cooling of the electrons.

This may point to a magnetic origin for the flares, which would also support a
transient jet scenario (\eg\ \citealt{Dexter14}). Furthermore, if we assume that
immediately prior to this flare the high-energy continuum is produced by thermal
Comptonization, following a simlar calculation to \cite{Merloni01} and taking $h$ to
be representative of the size-scale of the corona, we find that there is not enough
thermal energy stored in the corona to power the flare by many orders of magnitude,
which would also support a magnetic origin. While the spectrum during the peak is
also likely too hard for direct synchrotron emission from a jet, which would be
expected to give $\Gamma \sim 1.7$ in the X-ray band, but synchrotron-self-Compton
emission (\eg\ \citealt{Markoff05}) may be able to produce a high-energy continuum
this hard. 

% Indeed, sources with strong jets are known to be able to produce
% extremely hard X-ray spectra

The increase of roughly an order of magnitude in $E_{\rm{cut}}$ observed
across flare 4, from $\sim$50 to $\sim$500\,keV, would also appear to indicate that
significant energy is being injected into the X-ray emitting electron population
during this event, as $E_{\rm{cut}}$ is a proxy for the electron temperature
$T_{\rm{e}}$. \integral\ may have seen a similar evolution in the cutoff across one
of the bright flares observed during its coverage of this outburst (\citealt{Natalucci15}).
If the height of the source does increase across this flare, the change in gravitational
redshift experienced by the primary emission could contribute at least in part to the
difference seen in $E_{\rm{cut}}$, since this correction is not yet incorporated into the
\relxilllp\ model (\citealt{Niedzwiecki16}). However, in the most extreme scenario,
where \rin\ remains constant while $h$ varies (evolving from $\sim$2 to $\sim$20\,\rg),
the movement of the source height should only result in a factor of $\sim$2 change in
the observed cutoff energy (assuming no intrinsic variation). This is clearly insufficient
to explain the difference observed, and so we conclude that the intrinsic cutoff energy
does indeed increase across the flare.

Assuming the powerlaw emission is produced by Compton scattering at least during
the times both prior to and after the main flare, this implies either an increase in the
characteristic electron temperature if the particle distribution remains thermal, or
perhaps a transition to a more powerlaw-like (non-thermal) distribution that extends
up to significantly higher energies. With the spectral coverage of \nustar\ stopping at
79\,keV, it can be difficult to distinguish between these two scenarios for sufficiently
high electron temperatures in the thermal case, as the high-energy cutoff is shifted
out of the \nustar\ bandpass, resulting in the observation of a powerlaw spectrum
with little or no curvature. In turn, this results in a run-away effect in terms of the
measured $E_{\rm{cut}}$, owing to the fact that the cutoff powerlaw model is
constantly curving at all energies, while a thermal Comptonization continuum is more
powerlaw-like until it rolls over with a sharper cutoff (see the discussion in
\citealt{Fuerst16}), potentially explaining the fact that $E_{\rm{cut}}$ is often
consistent with the maximum value currently permitted by the \relxill\ models after
the flare. Nevertheless, the evolution in $E_{\rm{cut}}$ observed here provides a
good match to the jet model described in \cite{Markoff05}, in which electrons are
accelerated into a powerlaw distribution within a region $\sim$10--100\,\rg\ above
the jet's point of origin. Should the particle distribution instead be thermal both before
and after the peak of the flare, assuming that the size of the corona increases across
the flare (\ie\ it expands as either \rin\ or $h$ increase), then the evolution would be
similar to that expected for a corona being kept close to its catastrophic pair
production limit (see \citealt{Fabian15}, and references therein).

% The thermal energy of an electron plasma is given by $E_{\rm{th}} \simeq \pi \tau
% R^{2} k T_{\rm{e}} / \sigma_{\rm{T}}$, where $\tau$ is the plasma optical depth,
% $R$ is its radius, $k$ is Boltzmann's constant and $\sigma_{\rm{T}}$ is the
% Thomson scattering cross-section (\eg\ \citealt{Merloni01}). Assuming the
% powerlaw emission is produced by Compton scattering during the times both prior
% to and after the main flare, we can estimate $T_{\rm{e}}$ and $\tau$ from the form
% of this continuum. However, the relation between $E_{\rm{cut}}$ and $T_{\rm{e}}$
% becomes non-linear when $k T_{\rm{e}}$ is significantly larger than the \nustar\
% bandpass, as is potentially the case here, owing to the subtle differences in the
% high-energy curvature between a Comptonization spectrum and a powerlaw with
% an exponential cutoff (see discussion in \citealt{Fuerst16}). Therefore, in order to
% estimate these quantities, we replace the powerlaw continuum with a Comptonized
% spectrum, using the \comptt\ model (\citealt{comptt}), keeping all other parameters
% fixed at their best-fit values. Taking T2 and T7 as examples of the pre-flare and
% post-flare spectra, respectively, we find temperatures of $kT_{\rm{e,pre}} \sim
% 37$\,keV and $kT_{\rm{e,post}} \sim 185$\,keV and optical depths of
% $\tau_{\rm{pre}} \sim 0.5$ and $\tau_{\rm{post}} \sim 0.1$.

Finally, the geometric results from our reflection modelling are also likely consistent
with a jet scenario. We see evidence for either the disk truncating or the height of the
X-ray source increasing, both on average as the source flux decreases, and also
across one of the major flares individually. Although we cannot constrain the evolution
of the inner disk radius and the source height simultaneously, as variations in the two
produce similar results for the observed reflection spectrum (\eg\ \citealt{Fabian14},
hence our treatment of these two possibilities in isolation), as noted previously it is
quite possible that both of these quantities evolve. Indeed, if we repeat the time
resolved analysis of flare 4 presented in Section \ref{sec_flare4} forcing this to be the
case, linking the two with a simple linear relation and assuming that both evolve
simultaneously just for illustration $($[$h$/\rh] = 2[\rin/\risco]$)$, we again find the
same qualitative evolution seen in Figure \ref{fig_flare4res}, and the fits are as good
as the scenarios in which only one of \rin\ and $h$ is allowed to vary. In this scenario,
the magnitude of the changes in \rin\ and $h$ are both reduced in comparison to the
de-coupled scenarios discussed in Section \ref{sec_flare4}, with \rin\ evolving from
1--5\,\risco, and $h$ evolving from 2--10\,\rh. We note that, should the ejecta have
reached a significant outflow velocity, the reflection fraction would be reduced for a
given combination of $h$ and \rin\ (\eg\ \citealt{Beloborodov99}) resulting in these
quantities potentially being overestimated during the times after the flare, but again
the same qualitative evolution should be seen. Furthermore, acceleration up to
significant outflow velocities may not be expected so close to the black hole (see
Section \ref{sec_launch}).

In a flare associated with transient jet ejection, obviously if the jet is the source of
illumination then one naturally expects the height of the source to increase across
the flare. However, such ejection events may also be associated with an evacuation
of the inner disk, as the same instability that results in the ejection also results in
catastrophic accretion of the innermost portion of the disk (\citealt{Szuszkiewicz98,
Meier01}). \cite{Chen95} suggest that thin disk solutions should become unstable
above luminosities of $\sim$0.3\,\ledd, similar to the peak disk fluxes inferred here.
Evidence for such behaviour might be seen, for example, in GRS\,1915+105, where
radio ejections are also preceded by dips in X-ray intensity in some of the
oscillatory states exhibited by this source, during which the inner radius of the
accretion disk is inferred to increase (\eg\ \citealt{Pooley97, Mirabel98, KleinWolt02}),
though \cite{Rodriguez08} suggest that the ejections might actually be associated
with the post-dip flares observed in those cycles. Similar behaviour may also have
been seen in the radio galaxies 3C\,120 (\citealt{Marscher02, Chatterjee09,
Lohfink13}), and 3C\,111 (\citealt{Chatterjee11}), where radio ejections appear to be 
preceeded by X-ray dips. Therefore, both an increasing source height \textit{and} a
truncation of the inner accretion disk may be expected for transient ejection events,
consistent with the evolution seen in our analysis.

\subsubsection{Radio Monitoring}

A natural prediction of the jet scenario is that radio emission should be observed.
Throughout this recent outburst, \v404\ was frequently monitored by the Arcminute
Microkelvin Imager - Large Array (hereafter AMI; \citealt{AMI}), a compact array of
eight dishes operating in the 13-18 GHz frequency range. The full AMI campaign on
\v404\ will be presented in Fender et al. (in preparation; see also \citealt{Mooley15});
here we focus on the coverage that is simultaneous with our \nustar\ observation.
Flagging and calibration of the data were performed with the AMI REDUCE software
(\citealt{Perrott13}). The calibrated data were then imported into CASA and flux
densities of \v404\ were extracted by vector averaging over all baselines; the
absolute flux calibration uncertainty is $\sim$5\%.

\begin{figure}
\hspace*{-0.5cm}
\epsscale{1.16}
\plotone{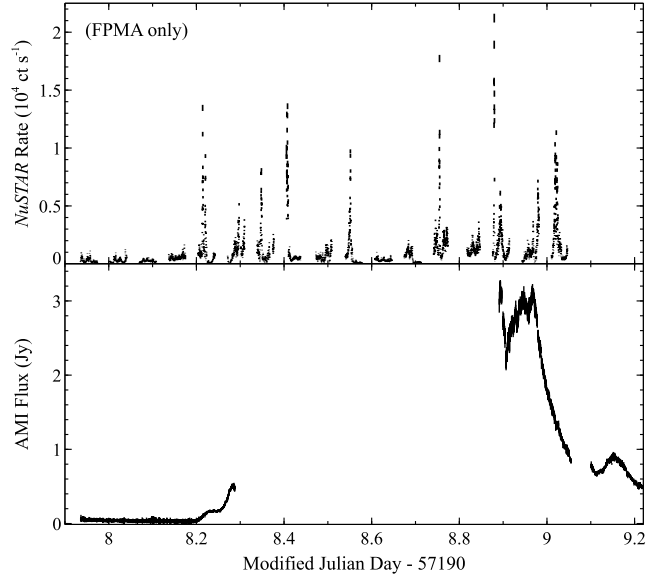}
\caption{A comparison of the \nustar\ lightcurve (top panel) and the radio monitoring
during this period from AMI (see text). Although there is a significant gap in the AMI
coverage owing to earth occultation, preventing a detailed analysis of the radio vs
X-ray behaviour, the overlapping coverage is sufficient to demonstrate the onset of
radio activity coincident with the strong flaring phase seen by \nustar.}
\vspace{0.3cm}
\label{fig_radio}
\end{figure}

A comparison of the \nustar\ and AMI lightcurves is shown in Figure \ref{fig_radio}.
Unfortunately, owing to occultation by the earth, the majority of the flaring period
observed by \nustar\ does not have simultaneous AMI coverage which, in
combination with the frequent earth-occultations experienced by \nustar, prevents
any detailed analysis attempting to search for radio responses to specific X-ray
flares. However, there is AMI coverage right at the beginning of this period, and
towards the end of the \nustar\ observation. These short periods of overlap do
clearly show that radio activity commences as the flaring phase of the \nustar\
observation begins, which then appears to persist throughout. The coincidence of
this radio activity further supports our suggestion that the major flares seen by
\nustar\ represent jet ejection events.

\subsubsection{Transient Jet Launching}
\label{sec_launch}

One of the most popular theoretical mechanisms for launching jets is that they are
powered by the spin of the central black hole (\citealt{BZ77}). The accreting black
hole system also may power a Blandford-Payne-type jet (\citealt{BP82}) powered
instead by the rotation of the accretion disk. It has been suggested that there is
observational evidence for a correlation between black hole spin and jet power
(taking the peak radio flux as a proxy for jet power) for the transient jet ejections
seen from other BHBs at high luminosities (\citealt{Narayan12, Steiner13}), as
expected for the \cite{BZ77} mechanism. However, this is still rather controversial
(\citealt{Russell13}).

If we are correct and these flares do represent jet ejections in which the jet is the
source of illumination for the disk, then our reflection analysis suggests that these
jets are launched from very close to the black hole (as close as a few \rg).
The size-scales inferred here are broadly comparable to the size-scale inferred
for the base of the jet in M87 (\citealt{Doeleman12, Asada12, Nakamura13}),
although this is a low-Eddington system that is likely analogous to the persistent
jets seen in the low/hard state of BHBs, rather than the high-Eddington transient
ejections potentially observed here. One of the other key results from the M87
system is that the acceleration of the outflowing plasma occurs gradually as the
distance from the black hole increases; jets do not seem to be immediately
launched with relativistic velocities (\citealt{Nakamura13}). This is an important
point, as it means that the emission from the regions of the jet close to the black
hole is unlikely to be heavily beamed, and can therefore illuminate the disk.

As a further point of interest, \cite{Koljonen15} present evidence for a relation
between the photon index of the high-energy X-ray continuum and the frequency at
which the low-energy synchrotron spectrum from the jet breaks from optically thick
to optically thin emission for a sample of accreting black holes, consisting of both
Galactic BHBs and AGN. This has been derived primarily from data obtained in the
low-Eddington jet regime (including low-Eddington observations of \v404). However,
should these high-Eddington ejections adhere to the same relation, the photon
indices of $\Gamma \sim 1.1$ seen during the peak of flare 4 would imply a break
frequency of $\sim$10$^{16}$\,Hz at this time. Unfortunately independent
observational constraints on the jet break are not available for this epoch, owing to
the lack of simultaneous radio--UV coverage. Nevertheless, should this be correct,
this would be among the highest break frequencies inferred among the sample
utilized by \cite{Koljonen15}, and would provide further, albeit indirect evidence that
the key jet activity in this case occurs very close to the black hole. Indeed, for the jet
model discussed in \cite{Markoff05}, a break frequency of $\sim$10$^{16}$\,Hz would
imply a height for the initial zone of particle acceleration of only a few \rg\ above
the base of the jet, which would be consistent with the geometric evolution across
this flare inferred from the reflection fits presented here.

While not a proof that these ejections are powered by black hole spin, the
size-scales inferred here do at least meet one of the expectations for the \cite{BZ77}
mechanism, that the jets should originate from regions very close to the black hole.
In addition, our work suggests that \v404\ hosts a rapidly rotating black hole (see
below), such that it is likely that there would be significant rotational energy for the
jets to tap into. However, we are not able to make any further assessment with
regards to the correlations presented by \cite{Narayan12} and \cite{Steiner13} with
these data, as it is highly plausible that the available radio coverage missed the peak
flux (Figure \ref{fig_radio}). The large gap in coverage also means we are not able to
reliably estimate the total energy of the radio flare, suggested by \cite{Fender10} and
\cite{Russell13} as an alternative proxy for jet power. The other major possibility, that
the jets are primarily powered by the disk rather than the black hole (\citealt{BP82}),
is also compatible with our results. In this scenario, the implied size-scales would
require that the jets be powered in the very innermost regions of the accretion disk.

\subsection{Black Hole Spin}
\label{sec_spin}

Through our investigation of the inner accretion geometry, we are also able to place
constraints on the spin of the black hole in \v404. Our initial modeling of the
flux-resolved spectra provided some indication that the black hole spin is high, owing
to the strong disk reflection inferred ($R_{\rm{disk}} \sim 3$). This requires strong
gravitational lightbending, which in turn requires a high black hole spin, such that the
disk can extend very close to the black hole and subtend a large solid angle as seen
by the illuminating X-ray source (\citealt{lightbending, Dauser13}). Evidence for strong 
gravitational lightbending has previously been observed in a wide variety of active
galactic nuclei (\eg\ \citealt{Zoghbi08, Fabian12, Parker14mrk, Reis14nat, Reynolds14,
Chiang15iras, Lanzuisi16}), but also in other Galactic BHBs (\eg\ \citealt{Rossi05,
Reis13lb}). Furthermore, as noted previously, potential evidence for strong reflection
has also been seen during flares seen by the \integral\ coverage of this outburst from
\v404 (\citealt{Roques15, Natalucci15}).

A high spin is supported by our flux-resolved analysis with a self-consistent lamppost
geometry. There is some complexity in the results obtained for the two scenarios
considered with the pure lamppost reflection model (varying the inner radius of the disk
while holding the height of the X-ray source constant, and vice versa; Models 2 and 3,
respectively), with similarly good fits obtained with high and more moderate spin
solutions in both cases. However, we obtain a significant improvement in the global fit
with the inclusion of a contribution from thermal disk emission at the highest fluxes in
addition to the lamppost component (Model 4); this is our best-fit model for the
flux-resolved data. In this case, a high spin is unambiguously preferred: $a^* > 0.82$
(see Figure \ref{fig_spin}, top panel).

In addition, a high spin is also supported by our flare-resolved analysis, focusing on
the peaks of the six most extreme flares observed. While this analysis utilizes much
less total exposure than our flux-resolved analysis, it has the advantage of relying on
much less averaging of different spectra (see Figure \ref{fig_allflares}). The pure
lamppost model strongly requires a high spin (Model 5), but we again see a significant
improvement in the fit with the inclusion of a thermal disk component (Model 6); this
is our best-fit model for the flare-resolved data. In this case, we see a strong
degeneracy between the black hole spin and the inclination of the inner accretion disk,
resulting in high- and moderate-spin solutions again providing similarly good fits.
The best-fit inclination, $i_{\rm{disk}} \sim 52$\deg, which corresponds to the high-spin
solution, is in good agreement with the range inferred for the orbital plane of the binary
system, $i_{\rm{orb}} \sim 50$--75\deg; \eg\ \citealt{Shahbaz94, Khargharia10}). If we
require the inclination to be in this range (Model 6i), then the spin is again strongly
required to be high: $a^* > 0.98$ (see Figure \ref{fig_spin}, bottom panel).
Taking a more conservative 99\% confidence level, the spin constraint expands to $a^*
> 0.92$. Given the lower degree of time-averaging of different spectral `states' in the
data analysed\footnote{While this does formally still occur to some minor
degree, this does not appear to have any significant effect on the results obtained. The
periods contributing to the Flare 4 spectrum considered in Section \ref{sec_flares} and
shown in Figure \ref{fig_allflares} are shaded blue in Figure \ref{fig_flare4spec}. These
are drawn from periods T2--6 shown in Figure \ref{fig_flare4res}, during which \v404\
does show some spectral variations (T2 is notably different to T3--6). However, if we
sum the data just from periods T2--5, where the observed spectra are all very similar,
the resulting spectra are practically identical to the Flare 4 spectra from Section
\ref{sec_flares}.} and the good agreement with the orbital inclination, we consider this
to be the most robust spin constraint derived from any of our models.

The quantitative constraints on the black hole spin discussed here are the statistical
parameter constraints obtained through our spectral modeling. There are additional
systematic errors associated with the assumptions inherent to the models used here
which are likely significant, but difficult to robustly quantify. One issue common to any
attempt to constrain black hole spin is the assumption that the accretion disk truncates
quickly at the ISCO, and that no significant emission should be observed from within
this radius. Numerical simulations suggest that, for thin disks, this is a reasonable
assumption (\eg\ \citealt{Shafee08, Reynolds08}), and that any additional uncertainty
should be small ($\sim$a few percent), particularly for rapidly rotating black holes.

For the particular case of \v404\ considered here, given the extreme luminosities
reached during the flares it is worth considering whether the assumption of a thin
disk is reasonable. Standard accretion theory predicts that as the accretion flow
becomes more luminous, its scale-height should start to increase as vertical support
from radiation pressure becomes more prominent (\eg\ \citealt{Shakura73}). Indeed,
some thickness to the disk may be required in order for the disk to be able to anchor
the magnetic fields required for jet ejections (\eg\ \citealt{Meier01, Tchekhovskoy12}).
This is potentially important for both the issue of how quickly the disk truncates at the
ISCO, as thicker disks are more able to exhibit emission that `spills over' the ISCO
slightly (\eg\ \citealt{Reynolds08}), and also for the self-consistent lamppost reflection
models, which calculate the expected reflection contribution assuming a thin disk
geometry (\citealt{Dauser13, relxill}).

In section \ref{sec_flares}, we estimated the peak disk luminosities to be $L_{\rm{disk}}
\sim 0.3-0.5$\,\ledd. Typically, the high-energy powerlaw emission from Galactic BHBs
is assumed to arise from Compton up-scattering of disk photons, and so the intrinsic
disk luminosities would have to be further corrected for the flux lost into the powerlaw
component (\eg\ \citealt{Steiner09}). This may well be the case at times outside of the
flare peaks. However, as noted above, during the flares the powerlaw emission is likely
too hard to originate via Compton scattering of disk photons. If we are correct about the
magnetic/jet ejection nature of these flares, then we should be able to take the peak disk
fluxes at roughly face value. Therefore, we take $L_{\rm{disk}}$/\ledd\ $\lesssim$ 0.5.
For the calculations of the expected disk structure presented in \cite{McClintock06}, this
would correspond to a maximum scale height of $h_{\rm{D}}/r_{\rm{D}} \lesssim 0.2$, or
equivalently a half-opening angle for the inner disk of $\lesssim$10\deg. This is unlikely
to be large enough that our assumption of a thin disk would lead to large errors. Even if
we are incorrect and the high-energy continuum does arise through up-scattering of
disk photons, since photon number (rather than flux) is conserved, the peak intrinsic
disk fluxes would only have been $\sim$20\% larger, even accounting for the hard
X-ray flux bent away from the observer in our strong lightbending scenario. Indeed,
\cite{Straub11} find that the X-ray spectrum of LMC X-3 is still fairly well described by a
thin disk model up to luminosities of $L_{\rm{disk}} \sim 0.6$\,\ledd. Furthermore, while
the flare peaks are extreme, the majority of the good exposure obtained naturally
covers lower fluxes, during which the thin disk approximation should be even more
reliable in terms of the reflection modeling. We therefore expect that, while there may
be some mild deviation from the thin disk approximation during the peaks of the flares
that could serve to relax the constraints on the spin slightly, this is unlikely to result in
major errors, and our conclusion that \v404\ hosts a rapidly rotating black hole is likely
robust to such issues.

\section{Conclusions}
\label{sec_conc}

The behaviour exhibited by \v404\ during its recent 2015 outburst is highly complex.
Our \nustar\ observation obtained during the height of this outburst activity revealed
extreme variability, both in terms of the observed flux and also the spectral properties
of \v404. In part, these variations are driven by strong and variable line-of-sight
absorption, as seen in previous outbursts from this source. However, strong flares
reaching $\sim$Eddington in the \nustar\ bandpass are also observed, during which
the central source appears to be relatively unobscured. These flares instead show
clear evidence for a strong contribution from relativistic reflection, providing a means
to probe the geometry of the innermost accretion flow. We argue these flares
represent transient jet ejection events, during which the ejected plasma is
the source of illumination for the accretion disk. This is based on the combination of
their observed properties, analogy with other Galactic BHBs, and also the
simultaneous onset of radio activity with the period of intense X-ray flaring observed.
If we are correct, then our modeling of the relativistic reflection with a lamppost
approximation implies that these jets are launched in very close proximity to the black
hole (within a few \rg), consistent with expectations for jet launching models that tap
either the spin of the central black hole, or rotation of the very innermost accretion
disk. In addition, our analysis allows us to place constraints on the black hole spin.
Although there are some quantitative differences between the different models
constructed, we consider our most robust spin constraint to be $a^* > 0.92$ (99\%
statistical uncertainty only). To the best of our knowledge, this is the first spin
constraint for \v404.

\section*{ACKNOWLEDGEMENTS}

The authors would like to thank the anonymous reviewer for their suggestions which
helped to improve the manuscript. DJW, PG and MJM acknowledge support from
STFC Ernest Rutherford fellowships (grant ST/J003697/2). KPM acknowledges
support from the Hintze Foundation. ALK acknowledges support from NASA through
an Einstein Postdoctoral Fellowship (grant number PF4-150125) awarded by the
Chandra X-ray Center, operated by the Smithsonian Astrophysical Observatory for
NASA under contract NAS8-03060. ACF acknowledges support from ERC Advanced
Grant 340442. LN wishes to acknowledge the Italian Space Agency (ASI) for
Financial support by ASI/INAF grant I/037/12/0-011/13 This research has made use of
data obtained with \nustar, a project led by Caltech, funded by NASA and managed by
NASA/JPL, and has utilized the \nustardas\ software  package, jointly developed by
the ASDC (Italy) and Caltech (USA). This research has also made use of data from
AMI, which is supported by the ERC, and we thank the AMI staff for scheduling
these radio observations.

{\it Facilities:} \facility{NuSTAR}, \facility{AMI}

\appendix

\section{A. Lower Spin Solutions}
\label{app_lowspin}

As discussed in the main text, for a number of the models presented in Sections
\ref{sec_flux} and \ref{sec_flares} we find the $\Delta\chi^{2}$ curves for the black hole
spin to show two similarly good solutions. Specifically, this is the case for our
flux-resolved analysis prior to the inclusion of an accretion disk component (Models 2
and 3, Section \ref{sec_flux}), and our flare-resolved analysis when the disk component
is included (Model 6, Section \ref{sec_flares}). Based on our flare-resolved analysis,
which minimizes the effects of averaging over different spectral forms, we favour the
high-spin case, as it is the solution that gives the best agreement between the inferred
disk inclination and the known orbital inclination. We therefore present the results for
the high-spin solutions for these models in the main manuscript. However, for
completeness, here we present the parameter constraints for the lower of the two spin
solutions found for Models 2, 3 and 6 (Tables \ref{tab_param_app} and
\ref{tab_flares_app}). As stated in the text, where these solutions are not the global
best fit, the errors are calculated as $\Delta\chi^{2} = 2.71$ around the local minimum.

\begin{table*}
  \caption{Results for the lower-spin solutions for Models 2 and 3 (flux-resolved analysis).}
  \vspace{-0.25cm}
\begin{center}
\begin{tabular}{c c c c c c c c c}
\hline
\hline
\\[-0.1cm]
Model Component & \multicolumn{2}{c}{Parameter} & Global & \multicolumn{5}{c}{Flux Level} \\
\\[-0.15cm]
& & & & F1 & F2 & F3 & F4 & F5 \\
\\[-0.2cm]
\hline
\hline
\\[-0.1cm]
\multicolumn{9}{c}{Model 2: truncating disk, static corona} \\
\\[-0.1cm]
\relxilllp\ & $\Gamma$ & & & $1.42^{+0.01}_{-0.05}$ & $1.44^{+0.01}_{-0.02}$ & $1.40\pm0.01$ & $1.37^{+0.01}_{-0.02}$ & $1.37\pm0.01$ \\
\\[-0.2cm]
& $E_{\rm{cut}}$ & [keV] & & $>620$\tmark[a] & $330\pm60$ & $190 \pm 10$ & $126^{+8}_{-6}$ & $92^{+4}_{-2}$ \\
\\[-0.2cm]
& $a^*$ & & $0.82^{+0.02}_{-0.07}$ \\
\\[-0.2cm]
& $i$ & [\deg] & $34\pm1$ \\
\\[-0.2cm]
& $h$ & \rh\ & $<2.1$ \\
\\[-0.2cm]
& $A_{\rm{Fe}}$ & [solar] & $3.0\pm0.1$ \\
\\[-0.2cm]
& \rin\ & \risco\ & & $1.7 \pm 0.2$ & $1.5 \pm 0.1$ & $1.3\pm0.1$ & $<1.2$ & 1 (fixed) \\
\\[-0.2cm]
& $R_{\rm{disk}}$\tmark[b] & & & 1.1 & 1.4 & 1.5 & 1.7 & 2.0 \\
\\[-0.2cm]
& Norm & & & $0.51 \pm 0.05$ & $0.64^{+0.25}_{-0.06}$ & $0.88^{+0.12}_{-0.04}$ & $1.27^{+0.44}_{-0.09}$ & $2.15^{+0.06}_{-0.08}$ \\
\\[-0.2cm]
\hline
\\[-0.1cm]
\chisq/DoF & & & 10657/10313 & \\
\\[-0.2cm]
\hline
\hline
\\[-0.1cm]
\multicolumn{9}{c}{Model 3: stable disk, dynamic corona} \\
\\[-0.1cm]
\relxilllp\ & $\Gamma$ & & & $1.38^{+0.05}_{-0.03}$ & $1.43^{+0.02}_{-0.01}$ & $1.39^{+0.02}_{-0.01}$ & $1.37 \pm 0.01$ & $1.38 \pm 0.01$ \\
\\[-0.2cm]
& $E_{\rm{cut}}$ & [keV] & & $>510$ & $320^{+30}_{-40}$ & $190 \pm 10$ & $126^{+5}_{-7}$ & $94 \pm 3$ \\
\\[-0.2cm]
& $a^*$ & & $0.64^{+0.05}_{-0.03}$ \\
\\[-0.2cm]
& $i$ & [\deg] & $31 \pm 1$ \\
\\[-0.2cm]
& $h$ & \rh\ & & $2.9^{+1.8}_{-0.4}$ & $2.8^{+0.6}_{-0.3}$ & $2.3 \pm 0.2$ & $<2.1$ & $<2.2$ \\
\\[-0.2cm]
& $A_{\rm{Fe}}$ & [solar] & $2.95^{+0.05}_{-0.06}$ \\
\\[-0.2cm]
& $R_{\rm{disk}}$\tmark[b] & & & 1.5 & 1.5 & 1.6 & 1.6 & 1.6 \\
\\[-0.2cm]
& Norm & & & $0.23^{+0.07}_{-0.04}$ & $0.32^{+0.04}_{-0.08}$ & $0.57^{+0.13}_{-0.02}$ & $1.08^{+0.10}_{-0.12}$ & $1.85^{+0.14}_{-0.35}$ \\
\\[-0.2cm]
\hline
\\[-0.1cm]
\chisq/DoF & & & 10674/10313 & \\
\\[-0.2cm]
\hline
\hline
\end{tabular}
\vspace{-0.2cm}
\label{tab_param_app}
\end{center}
$^a$ $E_{\rm{cut}}$ is constrained to be $\leq$1000\,keV following \cite{Garcia15}. \\
$^b$ For these models, $R_{\rm{disk}}$ is calculated self-consistently in the
lamppost geometry from $a^*$, $h$ and \rin. As it is not a free
parameter, errors are not estimated.
\vspace{0.4cm}
\end{table*}

\begin{table*}
  \caption{Results for the lower-spin solution for Models 6 (flare-resolved analysis).}
  \vspace{-0.25cm}
\begin{center}
\begin{tabular}{c c c c c c c c c c}
\hline
\hline
\\[-0.1cm]
Model Component & \multicolumn{2}{c}{Parameter} & Global & \multicolumn{6}{c}{Flare} \\
\\[-0.15cm]
& & & & 1 & 2 & 3 & 4 & 5 & 6 \\
\\[-0.2cm]
\hline
\hline
\\[-0.1cm]
\multicolumn{10}{c}{Model 6: lamppost with disk emission} \\
\\[-0.1cm]
\tbabs$_{\rm{src}}$ & \nh\ & [$10^{22}$ cm$^{-2}$] & & $0.9^{+0.9}_{-0.6}$ & $1.7^{+0.5}_{-0.9}$ & $<2.3$ & $<1.2$ & $1.9^{+1.2}_{-1.0}$ & $2.0 \pm 0.7$ \\
\\[-0.2cm]
\diskbb\ & $T_{\rm{in}}$ & [keV] & $0.50^{+0.08}_{-0.04}$ \\
\\[-0.2cm]
& Norm & [$10^5$] & & $<0.3$ & $1.2^{+1.3}_{-0.7}$ & $1.0^{+1.6}_{-0.5}$ & $0.7^{+1.0}_{-0.2}$ & $1.2^{+1.8}_{-0.7}$ & $1.1^{+1.4}_{-0.6}$ \\
\\[-0.2cm]
\relxilllp\ & $\Gamma$ & & & $<1.03$\tmark[b] & $1.29^{+0.04}_{-0.02}$ & $1.37 \pm 0.08$ & $1.37 \pm 0.07$ & $1.17^{+0.05}_{-0.08}$ & $1.23^{+0.04}_{-0.05}$ \\
\\[-0.2cm]
& $E_{\rm{cut}}$ & [keV] & & $40 \pm 2$ & $119^{+15}_{-14}$ & $37^{+5}_{-4}$ & $67^{+14}_{-11}$ & $50^{+2}_{-7}$ & $80^{+11}_{-6}$ \\
\\[-0.2cm]
& $a^*$ & & $0.45^{+0.42}_{-0.25}$ \\
\\[-0.2cm]
& $i$ & [\deg] & $<20$ \\
\\[-0.2cm]
& $h$ & \rh\ & & $2.9^{+2.7}_{-0.7}$ & $6.2^{+4.0}_{-3.2}$ & $11.1^{+21.2}_{-4.2}$ & $6.6^{+7.1}_{-2.1}$ & $7.6^{+32.6}_{-2.7}$ & $11.2^{+19.6}_{-5.8}$ \\
\\[-0.2cm]
& $\log\xi_{\rm{disk}}$ & $\log$[\ergcmps] & & $3.6 \pm 0.1$ & $4.0 \pm 0.1$ & $3.7 \pm 0.1$ & $3.8 \pm 0.1$ & $3.8 \pm 0.1$ & $4.0 \pm 0.1$ \\
\\[-0.2cm]
& $A_{\rm{Fe}}$ & [solar] & $5.0^{+0.3}_{-0.1}$ \\
\\[-0.2cm]
& $R_{\rm{disk}}$\tmark[a] & & & 1.4 & 1.2 & 1.1 & 1.2 & 1.2 & 1.1 \\
\\[-0.2cm]
& Norm & & & $0.9 \pm 0.4$ & $0.8^{+0.3}_{-0.1}$ & $0.4 \pm 0.1$ & $0.6^{+0.2}_{-0.1}$ & $0.6 \pm 0.1$ & $0.6^{+0.2}_{-0.1}$ \\
\\[-0.2cm]
\xstar$_{\rm{abs}}$ & $\log\xi$ & $\log$[\ergcmps] & $5.0^{+0.4}_{-0.3}$ \\
\\[-0.2cm]
& \nh\ & [$10^{21}$ cm$^{-2}$] & & $29^{+12}_{-7}$ & $<3$ & $<10$ & $<10$ & $<9$ & $<6$ \\
\\[-0.2cm]
\xillver\ & Norm & & & $0.20^{+0.04}_{-0.05}$ & $0.26^{+0.10}_{-0.05}$ & $0.13 \pm 0.07$ & $0.23^{+0.08}_{-0.09}$ & $0.33 \pm 0.06$ & $0.33^{+0.08}_{-0.04}$ \\
\\[-0.2cm]
\hline
\\[-0.15cm]
\chisq/DoF & & & 5906/5852 & \\
\\[-0.2cm]
\hline
\hline
\end{tabular}
\vspace{-0.2cm}
\label{tab_flares_app}
\end{center}
$^a$ For these models, $R_{\rm{disk}}$ is calculated self-consistently in the
lamppost geometry from $a^*$ and $h$. As it is not a free
parameter, errors are not estimated. \\
$^b$ The RELXILLLP model is only calculated for $\Gamma \geq 1$. \\
\vspace{0.5cm}
\end{table*}

\bibliographystyle{/Users/dwalton/papers/mnras}

\bibliography{/Users/dwalton/papers/references}

\label{lastpage}

\end{document}